\documentclass[trackchanges]{aastex701}
\usepackage{amsmath}

\usepackage{graphicx}
\usepackage{subcaption} 

\begin{document}

\title{Numerical Simulations of the Circularized Accretion Flow in Population III Star Tidal Disruption Events. I. The Accretion Flow and the Wind}

\author[orcid=0009-0008-9058-7023,gname=Yu-Heng,sname='Sheng']{Yu-Heng Sheng}
\affiliation{Astrophysics Division, Shanghai Astronomical Observatory, Chinese Academy of Sciences, 80 Nandan Road, Shanghai 200030, People’s Republic of China}
\affiliation{School of Astronomy and Space Sciences, University of Chinese Academy of Sciences, 19A Yuquan Road, Beijing 100049, China}
\email[]{shengyuheng@shao.ac.cn} 

\author[orcid=0000-0002-0427-520X,gname=De-Fu, sname='Bu']{De-Fu Bu} 
\affiliation{Shanghai Key Lab for Astrophysics, Shanghai Normal University, 100 Guilin Road, Shanghai 200234, China}
\email[show]{dfbu@shnu.edu.cn}

\author[gname=Xiao-Hong, sname='Yang']{Xiao-Hong Yang} 
\affiliation{Department of Physics, Chongqing University, Chongqing 400044, China }
\email[show]{yangxh@cqu.edu.cn}

\author[gname=Yi-ren, sname='Chang']{Yi-Ren Chang} 
\affiliation{Shanghai Key Lab for Astrophysics, Shanghai Normal University, 100 Guilin Road, Shanghai 200234, China}
\email[show]{changyiren@shnu.edu.cn}

\author[orcid=0000-0002-1908-0536, gname=Liang, sname='Chen']{Liang Chen}
\affiliation{Astrophysics Division, Shanghai Astronomical Observatory, Chinese Academy of Sciences, 80 Nandan Road, Shanghai 200030, People’s Republic of China}
\email[]{chenliang@shao.ac.cn}


\begin{abstract}
Tidal Disruption Events (TDEs) have recently been proposed as potential probes for Population III stars. However, the properties of the accretion flow and the wind from the Pop III star TDE system are not clear. By performing radiative hydrodynamic simulations, we study the 'circularized' accretion flow of the Pop III star TDE system. The masses of the black hole (BH) and the disrupted star are $10^6$ and $300$ solar masses, respectively. We focus on the properties of the wind. We find that the black hole accretion rate is highly super-Eddington. A strong wind is driven by radiation pressure. Due to the presence of a strong wind, only $25\%$--$35\%$ of the fallback debris is accreted by the BH. The remaining part is taken away by the wind. The kinetic power of the wind can be as high as $10^{46} {\rm \ erg \ s^{-1}}$. The properties of the wind obtained in this paper may be useful for understanding the radiation properties of Pop III star TDEs in the context of the wind 'reprocessing' model.  
\end{abstract}


\section{INTRODUCTION}

Stars in galactic nuclei may occasionally be scattered into orbits passing near the central supermassive black hole (SMBH). If the pericenter ($R_{\rm p}$) is equal to or smaller than the tidal disruption radius ($R_{\rm T}$) (\cite{1975Natur.254..295H}), the star can
be tidally disrupted, triggering a tidal disruption event (TDE; e.g., \cite{Rees1988}; \cite{Evans1989}). Approximately half of the stellar mass is unbound and ejected, while the remaining half is bound and falls back towards the SMBH. The gas fallback rate is predicted to decline with time roughly as $\dot M_{\rm fb} \propto t^{-5/3}$ (\cite{Evans1989};  \cite{Phinney1989}; \cite{Lodatoetal.2009}; \cite{Guillochon&Ramirez-Ruiz2013}).

For a typical TDE, in which a solar-type star is disrupted by a SMBH with mass $M_{\rm BH}=10^6$--$10^7 M_\odot$ (with $M_\odot$ being solar mass), the accretion flow is expected to have a size of $\sim 10^{13} {\rm cm}$. The effective temperature of the accretion flow is \(\sim\!\text{a few} \times 10^{5}\) K (see \cite{2020SSRv..216..114R} for reviews). This would imply peak emission in the soft X-ray band (\cite{Cannizzoetal.1990;}; \cite{Ulmer1999}; see \cite{2020SSRv..216..114R} for reviews). Observations in the 1990s indeed identified TDEs in soft X-ray bands (see \cite{Komossa2015} for a review). 

Over the past decade, many TDEs have been detected in the optical/ultraviolet (UV) band. These events indicated a significantly larger radiation size \( \sim 10^{14}\)–\(10^{16}\) cm (\cite{Hungetal2017}; \cite{vanVelzenet42al.2020}; \cite{Gezari2021}) and a much lower radiation temperature (\(\sim\!10^{4}\) K; e.g., \cite{Gezarietal.2012}; \cite{Arcavietal.2014)}). Two scenarios have been proposed to explain the optical/UV emission. For the first scenario, it is proposed that during the fallback of debris, shocks are produced in the collision process of streams (\cite{Guillochon&Ramirez-Ruiz2015}; \cite{Shiokawaetal.2015}), which dissipate energy and power the optical/UV emission (\cite{(Piranetal.2015}; \cite{JiangGuillochonLoeb2016}; \cite{Steinberg&Stone2024}; \cite{2025ApJ...979..235G}). The alternative "reprocessing" scenario suggests that the soft X-ray/extreme ultraviolet (EUV) photons generated close to the black hole are reprocessed into optical/UV bands by a geometrically extended debris envelope (\cite{Loeb&Ulmer1997}; \cite{Coughlin&Begelman2014}; \cite{Rothetal.2016}; \cite{Liuetal2017}; \cite{Liuetal.2021}; \cite{Metzger&Stone2017}; \cite{Metzger2022}; \cite{Weversetal.2022}), or by a wind (\cite{Strubbe&Quataert2009}; \cite{Lodato&Rossi2011}; \cite{Metzger&Stone2016}; \cite{Piro&Lu2020}; \cite{Uno&Maeda2020}; \cite{BU22}; \cite{Parkinsonetal.2022}; \cite{MageshwaranShawBhattacharyya2023}; \cite{Curd&Narayan2019}; \cite{Dai2018}).

Recently, delayed radio emission of TDEs has been observed (\cite{2020SSRv..216...81A}; \cite{2021NatAs...5..491H}; \cite{2022ApJ...938...28C}; \cite{2022ApJ...925..143P}; \cite{2022MNRAS.511.5328G}; \cite{2022ApJ...933..176S}; \cite{2024ApJ...962L..18Z}). It is believed that this delayed radio emission may be caused by the interaction between the TDE winds and the surrounding medium. When the TDE winds move outward, they will collide with the smoothly distributed circum-nuclear medium (CNM), which will generate shocks, accelerate cosmic ray electrons and amplify magnetic fields. It is found that the delayed radio emission may be generated via the synchrotron radiation process in the wind-CNM collision process (\cite{2013ApJ...772...78B}; \cite{2021MNRAS.507.4196M,2024ApJ...971...49M}; \cite{2024ApJ...971..185C}; \cite{2025ApJ...988L..24H}). Very recently, it has been found that the TDE winds can collide with dense clouds surrounding the black hole, producing bow shocks. The cosmic ray electrons can be accelerated in the bow shocks. Recently, theoretical models of wind-cloud interaction have found that the radio emission of some TDEs can be well explained (\cite{Mou2022}; \cite{2023MNRAS.521.4180B}; \cite{2024ApJ...977...63L}; \cite{2025ApJ...979..109Z}).
 
Due to the great importance of TDE winds in explaining the observations, we recently performed simulations to study the detailed properties of the wind from a circularized accretion flow in a typical TDE with masses of black hole and disrupted star being $10^{6-7} M_\odot$ and $1\ M_\odot$, respectively (\cite{BU23}). We find that the wind can take away more than half of the fallback debris. The velocity of the wind can be as high as $0.7c$, with $c$ being the speed of light. The kinetic power can be well above $10^{44} {\rm erg \ s^{-1}}$. The properties of the wind from a circularized accretion flow in typical TDEs have also been studied by other groups (\cite{Curd&Narayan2019}; \cite{Dai2018}; \cite{Thomsen2022}). We note that winds from the self-colliding shock process between the fallback streams have also been studied recently (\cite{JiangGuillochonLoeb2016}; \cite{Bonnerot&Lu2020}). Recently, observations of TDEs have confirmed the presence of TDE winds via UV and X-ray spectra (\cite{2017ApJ...846..150Y} ; \cite{2018MNRAS.474.3593K} ; \cite{2015Natur.526..542M}; \cite{2020MNRAS.494.4914P};
\cite{2025ApJ...989L...9L}; \cite{2016ApJ...818L..32C}; \cite{2024SciA...10J8898P}; \cite{2023ApJ...954..170K}; \cite{2015ApJ...811...43L};
\cite{2023eas..conf.1878W};
\cite{2024ApJ...972..106X};
\cite{2014ApJ...780...44C};
\cite{2017ApJ...843..106B};
\cite{2018MNRAS.473.1130B}; 
\cite{2019ApJ...873...92B}; \cite{2019ApJ...879..119H};
\cite{2021ApJ...917....9H};
\cite{2023ApJ...951..134P}).

In this paper, we study the accretion flow and the wind in Pop III star TDEs. In the following, we introduce the motivation for this work in detail. Pop III stars form in the early Universe at redshift $z \sim 10$--$15$ (\cite{2002Sci...295...93A}; \cite{2002ApJ...564...23B}; \cite{2006ApJ...652....6Y}) with a mass in the range of $30M_\odot-300M_\odot$. Pop III stars are believed to have formed from the pristine gas a few hundred million years after the Big Bang. The newly formed Pop III stars are metal-free. Metals can be produced in the core of Pop III stars. Metals can be released into the intergalactic medium (IGM) via supernova explosions (\cite{2024ApJ...964...91C}). This process will shape the subsequent formation of higher-metallicity Pop II and Pop I stars. Therefore, directly detecting and studying the properties of Pop III stars is important to understand the generation of Pop II and Pop I stars. However, direct observations of Pop III stars are extremely difficult. Only a few Pop III-like star candidates have been identified (\cite{2020MNRAS.494L..81V}; \cite{2022Natur.603..815W}). \cite{ChowdhuryChang&Daietal.2024} proposed that Pop III stars can be indirectly detected by the Pop III star TDE process. They analytically studied the 'reprocessing' model for TDEs of the Pop III stars. They found that the wind of TDEs can reprocess the hard photons generated close to the black hole into optical/UV bands. Finally, a large fraction of the TDE photons are redshifted to infrared bands, which can be detected by JWST and the Nancy Grace Roman Space Telescope (Roman). We note that in the model of \cite{ChowdhuryChang&Daietal.2024}, the properties of the wind are artificially set. Their results may strongly depend on the assumption of the properties of the wind. Therefore, it is essential to study the properties of the accretion flow and the wind of Pop III star TDEs in detail by numerical simulations. One can use the properties of the wind obtained by our simulations to re-study the `reprocessing' model to obtain the radiation information of Pop III star TDEs. 

Besides that, according to \cite{ChowdhuryChang&Daietal.2024}, the debris fallback rate of Pop III star TDEs is several orders of magnitude higher than that of Pop I star TDEs. The increase of fallback rate increases the accretion rate. We expect that the characteristics of such a special accretion flow may differ from those of normal TDEs, and we are also interested in the behavior of accretion flows at such high accretion rates. For example, \cite{ChowdhuryChang&Daietal.2024} presented that for a $10^6M_\odot$ black hole, the peak debris fallback rate of Pop III star TDEs is $\dot M_{\rm fb, peak} \sim 10^{4-6} \dot M_{\rm Edd}$, where $\dot{M}_{\rm Edd} = \frac{L_{\rm Edd}}{\eta c^2}$ is the Eddington accretion rate, with $L_{\rm Edd}$ being the Eddington luminosity and $\eta$ is the radiative efficiency. We assume $\eta = 0.1$ when we calculate the Eddington accretion rate. For comparison, in Pop I star TDEs where $M_{\rm BH} = 10^6M_\odot$ and the disrupted star is a solar-type star, the peak debris fallback rate is $\dot M_{\rm fb, peak} \sim 100 \dot M_{\rm Edd}$. There have been many numerical simulations studying the super-Eddington accretion flow (e.g.,\cite{2005ApJ...628..368O}; \cite{Dai2018};\cite{Curd&Narayan2019};\cite{Curd&Narayan2023}; \cite{Thomsen2022}; \cite{BU23}). However, to our knowledge, there are no simulations that study the super-Eddington accretion flow with an accretion rate as high as $10^{4-6} \dot M_{\rm Edd}$. Therefore, from a theoretical point of view, it is of great importance to study the accretion system of Pop III star TDEs.

In this paper, we present numerical simulations to study the accretion flow and the wind of Pop III star TDEs. We focus on the detailed properties of the wind.
The structure of this paper is as follows: In Section 2 we describe the physical framework and computational methodology employed in our simulations. In Section 3 we present the results of our numerical simulations. We summarize and discuss our results in Section 4.

\section{NUMERICAL METHOD} 
We performed 2D axisymmetric hydrodynamic simulations using the PLUTO code (\cite{Mignone2007}), in which a radiative transfer module is incorporated. We solved the governing equations listed below in spherical polar coordinates ($r$, $\theta$, $\phi$) 

\begin{equation} 
\frac{\mathrm{d} \rho}{\mathrm{d} t} + \rho \nabla \cdot \mathbf{v} = 0
\end{equation}

\begin{equation} 
\rho \frac{\mathrm{d} \mathbf{v}}{\mathrm{d} t} = -\nabla p - \rho \nabla \psi + \nabla \cdot \mathbf{T} + \mathbf{f}_{\mathrm{rad}}
\end{equation}

\begin{equation} 
\rho \frac{\mathrm{d}}{\mathrm{d} t}\left(\frac{e}{\rho}\right) = -p \nabla \cdot \mathbf{v} - 4\pi\kappa B + c\kappa E_{\mathrm{rad}} + Q_{\mathrm{vis}}
\end{equation}

\begin{equation} 
\rho \frac{\mathrm{d}}{\mathrm{d} t}\left(\frac{E_{\mathrm{rad}}}{\rho}\right) = -\nabla \cdot \mathbf{F}_{0} - \nabla \mathbf{v} : \mathbf{P}_{0} + 4\pi\kappa B - c\kappa E_{\mathrm{rad}}
\end{equation}

Here, $\rho$, $\mathbf{v}$, $p$, $e$, and $E_{\mathrm{rad}}$ denote the gas density, velocity, gas pressure, gas internal energy density, and radiation energy density, respectively. We employ an adiabatic equation of state $p = (\gamma - 1)e$, with $\gamma = 5/3$. The gravitational potential $\psi$ of the black hole is described by the pseudo-Newtonian potential $\psi = -GM_{\mathrm{BH}}/(r - R_{\rm s})$, where $G$, $M_{\mathrm{BH}}$, and $R_{\rm s}$ represent the gravitational constant, the black hole mass and the Schwarzschild radius, respectively. The stress tensor $\mathbf{T}$ is responsible for angular momentum transfer, $Q_{\mathrm{vis}}$ denotes the viscous heating rate, $\mathbf{F_0}$ denotes the radiative flux, $\mathbf{P_0}$ represents the radiation pressure tensor, $\kappa$ is the absorption opacity, and $\mathbf{f}_{\mathrm{rad}}$ is the radiation force. Finally, $B$ corresponds to the blackbody intensity.

For radiation transfer, we employ the absorption opacity $\kappa_{\mathrm{ab}} = \kappa_{\mathrm{ff}} + \kappa_{\mathrm{bf}}$, where the free-free absorption coefficient is given by $\kappa_{\mathrm{ff}} = 1.7 \times 10^{-25} T^{-7/2} (\rho / m_{\rm p})^2 \mathrm{cm^{-1}}$ and the bound-free absorption coefficient follows $\kappa_{\mathrm{bf}} = 4.8 \times 10^{-24} T^{-7/2} (\rho / m_{\rm p})^2 (Z / Z_{\odot}) \mathrm{cm^{-1}}$ (\cite{Hayashi1962}; \cite{Rybicki1979}). Here, $T$ and $m_p$ denote the gas temperature and the proton mass, respectively. $Z$ and $Z_{\odot}$ represent the metallicity and the solar metallicity, respectively. Previous works show that the critical metallicity of Pop III stars is $\sim 10^{-3}-10^{-5}Z_\odot$, above which the transition from Pop III stars to Pop II stars occurs (\cite{2001MNRAS.328..969B}; \cite{2002ApJ...571...30S}; \cite{2004ApJ...605..579Y}; \cite{2012ApJ...745...50W}; \cite{2018MNRAS.475.4396J}). Following \cite{ChowdhuryChang&Daietal.2024}, we choose two values of metallicity $Z = 10^{-9} Z_{\odot}$ and $Z = 10^{-5} Z_{\odot}$. The radiation force is defined as $\mathbf{f}_{\mathrm{rad}} = \frac{\chi}{c} \mathbf{F}_{0}$, where the total opacity is $\chi = \kappa_{\mathrm{ab}} + \kappa_{\mathrm{es}}$. $\kappa_{\mathrm{es}}$ is the electron scattering opacity expressed in the standard formulation $\kappa_{\mathrm{es}} = 0.34\rho \mathrm{\ cm^{-1}}$. Radiation transport is modeled via the flux-limited diffusion approximation (\cite{Levermore1981}), expressed as $\mathbf{F}_{0} = -\frac{c\lambda}{\chi} \nabla E_{\mathrm{rad}}$, where the flux limiter $\lambda$ is defined by $\lambda =\frac{2 + \Re} {6 + 3\Re + \Re^2}$ with $\Re = |\nabla E_{\mathrm{rad}}| / (\chi E_{\mathrm{rad}})$. 
The radiation pressure tensor is given by $\mathbf{P}_{0} = \mathbf{f} E_{\mathrm{rad}}$, where $\mathbf{f}$ denotes the Eddington tensor defined by $\mathbf{f} = \frac{1}{2}(1-f)\mathbf{I} + \frac{1}{2}(3f-1)\mathbf{n}\mathbf{n}$, with the scalar Eddington factor $f$ calculated as $f = \lambda + \lambda^{2}\Re^{2}$. $\mathbf{n}$ is given by $\mathbf{n} = \nabla E_{\mathrm{rad}} / |\nabla E_{\mathrm{rad}}|$.

For viscosity, only the $r\phi$ component of the viscous stress tensor is non-zero: $T_{r\phi} = \eta r \frac{\partial}{\partial r} \left( \frac{v_{\phi}}{r} \right)$. The dynamic viscosity coefficient $\eta = \alpha \frac{p + \lambda E_{\mathrm{rad}}}{v_{\phi}/r} $ follows the prescription of \cite{2005ApJ...628..368O} with $\alpha = 0.1$ in this work. The viscous heating term is then given by $Q_{\mathrm{vis}} = \eta \left[ r \frac{\partial}{\partial r} \left( \frac{v_{\phi}}{r} \right) \right]^{2}$.

We assume that the star follows a parabolic trajectory towards the black hole. Pop III stars can be completely disrupted with a penetration factor $\beta \equiv R_{\rm T} / R_{\rm p} \geq 1.85$ (\cite{Guillochon&Ramirez-Ruiz2013}). The tidal radius $R_{\rm T} = R_{\star} (M_{\mathrm{BH}} / M_{\star})^{1/3}$ \cite{1975Natur.254..295H}, where $R_{\star}$ and $M_{\star}$ are the radius and mass of the star, respectively. In this paper, we set $\beta$ to 1.85. We use the mass-radius relation for Pop III stars given by \cite{Bromm2001b}:
\begin{equation}
R_{\star} \simeq 0.7 R_{\odot} \left( \frac{M_{\star}}{M_{\odot}} \right)^{0.45} \left( \frac{Z}{10^{-9}} \right)^{0.09}.
\label{eq5}
\end{equation}
In this paper, we set the mass of the disrupted star to $M_\star = 300 M_\odot$.
The disrupted bound debris falls back to the pericenter. Subsequently, we assume that the debris can be rapidly circularized and forms an accretion disk with a size of $R_{\rm C} = 2R_{\rm p}$. The transition from metal-free Pop III stars to metal-poor Pop II stars may occur at critical metallicity $\sim$ $10^{-3}-10^{-5}$ solar metallicity (e.g., \cite{Bromm2001b}). Since no prior research has described the properties of such an extremely high accretion rate system, in order to probe the metallicity dependence, as done in \cite{ ChowdhuryChang&Daietal.2024} , we choose two values for the metallicity $10^{-9}$ and $10^{-5}$ solar metallicity. In the simulations, we inject gas into the system at $R_{\rm C}=2R_{\rm p}$ with Keplerian angular velocity. The specific injection radius is $R_{\rm C} = 34.68 R_{\rm s}$ for the star with metallicity $Z = 10^{-9} Z_{\odot}$ and is $R_{\rm C} = 79.44 R_{\rm s}$ for the star with metallicity $Z = 10^{-5} Z_{\odot}$. The injected gas has zero velocity in both radial ($r$) and polar ($\theta$) directions.

When a star undergoes tidal disruption, if the mass distribution as a function of the binding energy is approximately flat, the debris returns to the system at a fallback rate $\dot{M}_{\mathrm{fb}} \propto t^{-5/3}$ (\cite{Rees1988}). In our simulations, the mass injection rate follows the same scaling:
\begin{equation}
\dot{M}_{\mathrm{inject}} = \dot{M}_{\mathrm{fb}} = \dot{M}_{\mathrm{fb,peak}} \left(1 + \frac{t}{t_{\mathrm{fb}}}\right)^{-5/3}
\label{eq6}
\end{equation}
where we adopt the characteristic fitting formulas from the high-precision hydrodynamics simulations conducted by \cite{Guillochon&Ramirez-Ruiz2013}:
\begin{equation}
t_{\mathrm{fb}} = B_{4/3} M_{6}^{1/2} m_{\star}^{-1} r_{\star}^{3/2} \, \mathrm{yr}
\end{equation}
\begin{equation}
\dot{M}_{\mathrm{fb,peak}} = A_{4/3} M_{6}^{-1/2} m_{\star}^{2} r_{\star}^{-3/2} \, M_{\odot} \, \mathrm{yr}^{-1}
\label{eq.fbpeak}
\end{equation}
with dimensionless parameters defined as $M_6 \equiv M_{\mathrm{BH}} / 10^{6} M_{\odot}$, $m_{\star} \equiv M_{\star} / M_{\odot}$, and $r_{\star} \equiv R_{\star} / R_{\odot}$. For radiation-dominated massive stars, the fitting coefficients are as follows:
\begin{equation}
B_{4/3} = \frac{ -0.38670 + 0.57291 \sqrt{\beta} - 0.31231 \beta }{ 1 - 1.2744 \sqrt{\beta} - 0.90053 \beta }
\end{equation}
\begin{equation}
A_{4/3} = \exp \left[ \frac{ 27.261 - 27.516 \beta + 3.8716 \beta^{2} }{ 1 - 3.2605 \beta - 1.3865 \beta^{2} } \right]
\end{equation}
Consistent with the methodology of \cite{ChowdhuryChang&Daietal.2024}, and using the mass-radius relation for Pop III stars, we use the following simplified expressions:

\begin{equation}
t_{\mathrm{fb}} \simeq 9 \left( \frac{B_{4/3}}{0.08} \right) \left( \frac{m_{\star}}{300} \right)^{-0.3} \left( \frac{Z}{10^{-5}} \right)^{0.1} M_{6}^{0.5} \, \mathrm{days}
\label{eq11}
\end{equation}

\begin{equation}
\dot{M}_{\mathrm{fb,peak}} \simeq 2.7 \times 10^{3} \left( \frac{A_{4/3}}{3} \right) \left( \frac{m_{\star}}{300} \right)^{1.3} \left( \frac{Z}{10^{-5}} \right)^{-0.1} M_{6}^{-0.5} \, M_{\odot} \, \mathrm{yr}^{-1}
\label{eq12}
\end{equation}

The computational domain spans radially from $2R_{\rm s}$ to $10^5 R_{\rm s}$ and in the angular direction from $\theta = 0$ to $\pi/2$. We employ a total of 768 grid points in the radial direction, distributed across four distinct regions with varying resolution: in the $2R_{\rm s} \leq r \leq 3R_{\rm s}$ region, we set 16 uniformly spaced grid points; in the $3R_{\rm s} \leq r \leq 50R_{\rm s}$ region, we set 240 logarithmically spaced grid points; in the $50R_{\rm s} \leq r \leq 200R_{\rm s}$ region, we set 128 logarithmically spaced grid points, and in the $200R_{\rm s} \leq r \leq 10^5 R_{\rm s}$ region, we set 384 logarithmically spaced grid points. In the angular direction, we implement 128 non-uniformly spaced grids with enhanced resolution near the disk mid-plane ($\theta = \pi/2$). Boundary conditions are specified as follows: Outflow conditions are applied at both inner ($r = 2R_{\rm s}$) and outer ($r = 10^5 R_{\rm s}$) radial boundaries; axisymmetric conditions are applied at the polar axis ($\theta = 0$), and reflective boundary conditions are applied at the equatorial plane ($\theta = \pi/2$), which enforces symmetry about the mid-plane.

\section{RESULTS}
In this section, we first present detailed simulation results for the $300 M_{\odot}$ star with metallicity $Z = 10^{-9} Z_{\odot}$ (hereafter referred to as model M300-9), followed by comparative results for another model. The black hole accretion rate $\dot{M}$ is computed at the inner boundary  of the simulation ($r = 2R_{\rm s}$) using:
\[
\dot{M} = 2 \times 2\pi \left(2 R_{\mathrm{s}}\right)^{2} \int_{0}^{\pi /2}\rho\min\left(v_{\mathrm{r}}, 0\right)\sin\theta\mathrm{d}\theta
\]
This formulation includes the factor of 2 to account for the simulation covering only the region above the mid-plane.

We now describe the method for analyzing wind flux. The accretion flow is quite turbulent. Therefore, the gas with $v_{\rm r} > 0$ may not represent the actual wind. It may be the outward moving portion of a turbulent eddy.  Following \cite{Curd&Narayan2023}, we define the wind using dual criteria: the Bernoulli parameter $\mathrm{Be} > 0$ and the positive radial velocity ($v_{\rm r} > 0$). 
As described in \cite{Curd&Narayan2023} (also referencing \cite{Yoshioka2022}), the Bernoulli parameter is calculated as follows:
\begin{equation}
\mathrm{Be} = 
\begin{cases} 
\frac{1}{2} v^{2} +  \frac{\gamma e}{\rho} +  \frac{{\gamma_{\mathrm{rad}}}E_{\mathrm{rad}}}{\rho} \left(1 - \tau_{\mathrm{es}}^{-1/2}\right) - \frac{GM_{\mathrm{BH}}}{r - R_{\rm s}}, & \tau_{\mathrm{es}} \geq 1  \\
\frac{1}{2} v^{2} +  \frac{\gamma e}{\rho} -  \frac{GM_{\mathrm{BH}}}{r - R_{\rm s}}, & \tau_{\mathrm{es}} < 1 
\end{cases}
\end{equation}
In our simulations, radiation is treated as a diffusive fluid with adiabatic index $\gamma_{\mathrm{rad}} = 4/3$. The Bernoulli parameter incorporates radiative contributions, with the electron scattering optical depth defined as $\tau(\theta, r) = \int_{ r}^{10^{5} R_{\mathrm{s}}} \rho \kappa_{\mathrm{es}} \mathrm{d} r^{\prime}$, representing the cumulative scattering experienced by radiation escaping from the grid point to the outer boundary. In optically thick regions ($\tau_{\mathrm{es}} > 1$), radiation couples strongly with gas, driving winds, and contributing positively to the Bernoulli parameter. In the optically thin region, radiation decouples from gas, and the Bernoulli parameter does not contain the contribution of radiation. The final mass flux and kinetic power of the winds are quantified through the following expressions:
\begin{equation}
\dot{M}_{\text{wind}} = 2 \times 2\pi r^{2} \int_{0}^{\pi/2} \max\left(\frac{\mathrm{Be}}{|\mathrm{Be}|}, 0\right)\rho\max\left(v_{\mathrm{r}}, 0\right)\sin\theta  \mathrm{d}\theta
\end{equation}
\begin{equation}
\dot{E}_{\text{wind}} = 2 \times 2\pi r^{2} \int_{0}^{\pi/2} \frac{1}{2}\max\left(\frac{\mathrm{Be}}{|\mathrm{Be}|}, 0\right) \rho  v_{\mathrm{r}}^{2} \max\left(v_{\mathrm{r}}, 0\right) \sin\theta  \mathrm{d}\theta
\end{equation}

\subsection{Wind in Pop III star TDEs with $M_\star = 300 M_\odot$ and metallicity $Z= 10^{-9} Z_\odot$ (Model m300-9)}
\begin{figure}[htbp]
\centering
\includegraphics[width=1.0\textwidth]{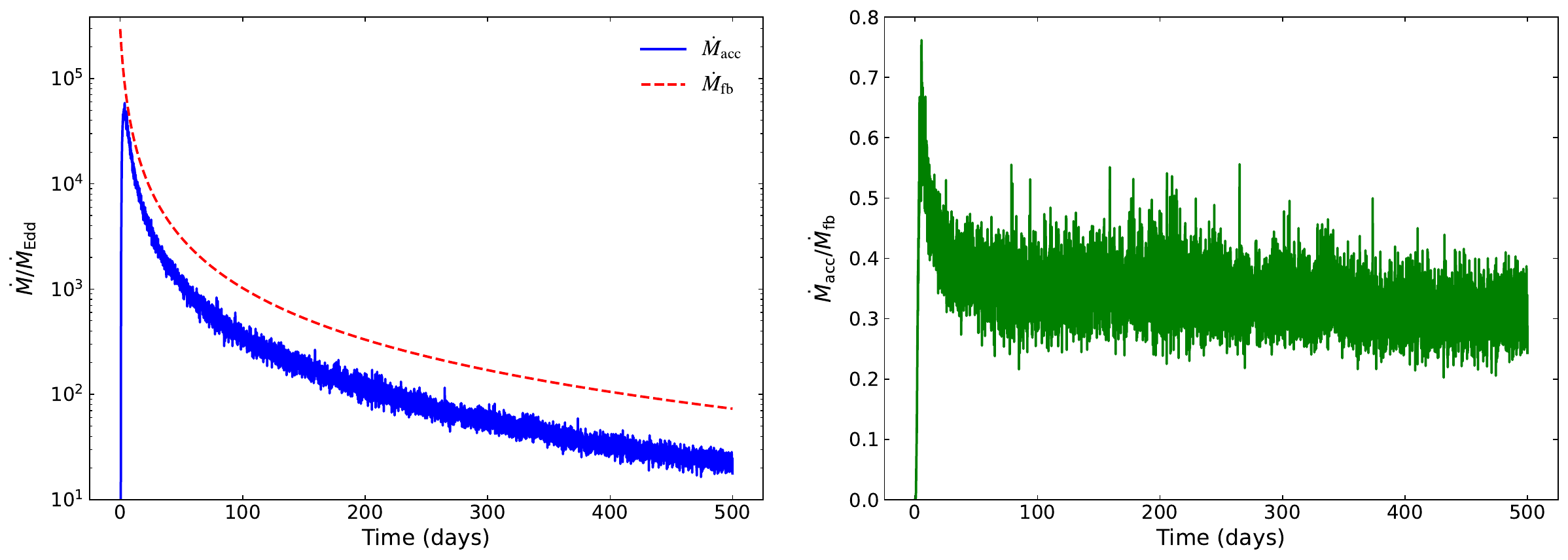}

\caption{Accretion rate for model M300-9. Left panel: time evolution of the black hole accretion rate (blue line) and stellar debris fallback rate 
(red dotted line) in Eddington units. Right panel: time evolution of black hole accretion in the unit of the stellar debris fallback rate.
}
\label{dot300-9}
\end{figure}

The left panel of Fig.~\ref{dot300-9} shows the time evolution of stellar debris fallback and black hole accretion rates. The right panel shows the ratio of the black hole mass accretion rate to the debris fallback rate. Due to the extremely high mass injection rate from the disruption of a Pop III star, the peak black hole mass accretion rate can reach $5.85\times10^4\dot{M}_{\rm Edd}$ at $ t = 3.7 \mathrm{\ days}$, which coincides with the dynamic timescale $(\frac{r}{v_{\rm r}}\simeq 2-3\ \rm{days})$ near the injection region. It is shorter than the viscous time scale (\(t_{vis}  \sim\frac{r^2}{\nu} \simeq 9\) days). This discrepancy arises because the inflow of gas is driven not only by angular momentum transfer but also by strong radiation pressure. 

The injected debris spreads outward due to the angular momentum transfer. The study by \cite{2014ApJ...784...87S} provides a theoretical framework for the evolution of TDE disks. In our simulation, we define the disk region based on the following criteria: the gas must have a specific angular momentum greater than $0.8\sqrt{GM_{\rm BH}r}$, and is rotationally supported, satisfying $\frac{v_{\rm r}^2+v_{\theta}^2}{v_{\phi}^2}\leq 0.05$. After approximately one dynamic timescale, the inner boundary of the accretion disk reached the simulation's inner boundary at 2$R_{\rm s}$. The outer boundary of the disk expands to approximately $ 400\rm R_{s}$, which is almost ten times the injection radius, at the end of the simulation. We then calculated the wind mass flux at 400 $\rm R_{\rm s}$ to study the wind launched from the disk and found that the wind mass flux is proportional to the accretion rate. The wind mass flux is consistently about twice the accretion rate, a result that coincides with the result shown in the right panel of Fig.~\ref{dot300-9}. Regarding the link between disk evolution and wind properties, we find that there is no relation between the disk's physical size and the wind mass flux. Considering that the wind mass flux is proportional to the accretion rate, we think that the wind is regulated by the accretion process itself, rather than being regulated by the instantaneous physical size of the disk.

The radiation pressure-dominated super-Eddington accretion flow is expected to drive powerful winds (e.g., \cite{Dai2018}; \cite{Curd&Narayan2019} ). For an extremely high accretion rate scenario, \cite{Cao&Gu2022} demonstrate that at mass supply rates of $\lesssim1000\,\dot{M}_{\mathrm{Edd}}$,
over $50\%$ of the gas becomes unbound in outflows. As seen in the right panel of Fig.~\ref{dot300-9}, after an initial rise near the peak, the ratio of the black hole accretion rate to the debris fallback rate is just around 35\%. Below, we quantitatively calculate how much of the fallback debris is accreted by the black hole over approximately 500 days:

\begin{equation}
f_{\mathrm{acc}} = \frac{M_{\text{accreted}}}{M_{\text{fallback}}} = \frac{\int_{0}^{499.89\,\mathrm{days}} \dot{M}_{\mathrm{acc}}  dt}{\int_{0}^{499.89\,\mathrm{days}} \dot{M}_{\mathrm{fb}}  dt} = 0.3499
\end{equation}
This indicates that only $\sim35\%$ of the stellar debris was accreted by the black hole. As a note, for a $10^6 M_\odot$ black hole disrupting a solar-type star, we find that $43\%$ of the fallback debris will be accreted by the black hole (\cite{BU23}).

\begin{figure}[htbp]
\centering
\includegraphics[width=1.0\textwidth]{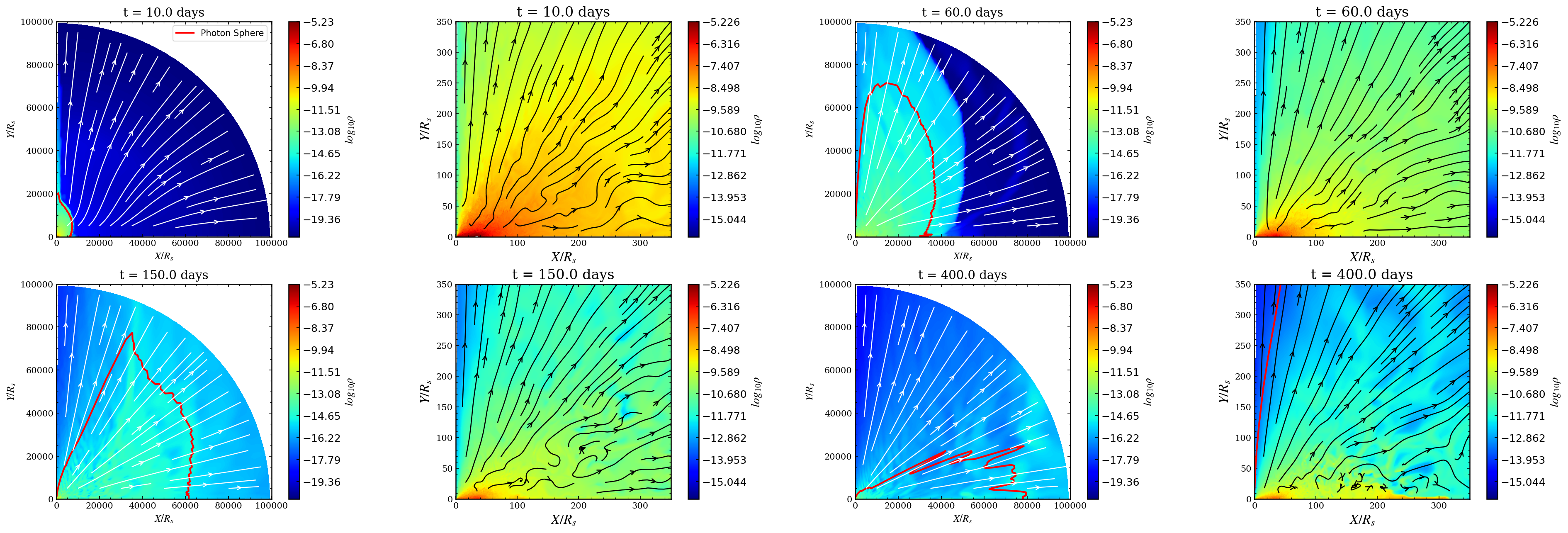}

\caption{
       Snapshots for gas density (colour scale) for model M300-9 with fluid velocity (streamlines). We illustrate the system state at 
    $t = 10.0\,\mathrm{days}$ (top-first and top-second panels), $t = 60.0\,\mathrm{days}$ (top-third and top-fourth panels), $t = 150.0\,\mathrm{days}$ (bottom-first and bottom-second panels), and $t = 400.0\,\mathrm{days}$ (bottom-third and bottom-fourth panels). For each time snapshot, we present two panels to show the properties of a zoom-out large domain and a zoom-in smaller domain of the system. The red lines in the top-first, top-third, bottom-first, bottom-third and bottom-fourth panels are the electron scattering photosphere.
}
\label{fig:evol300-9}
\end{figure}

\begin{figure}[htbp]
\centering
\includegraphics[width=0.9\textwidth]{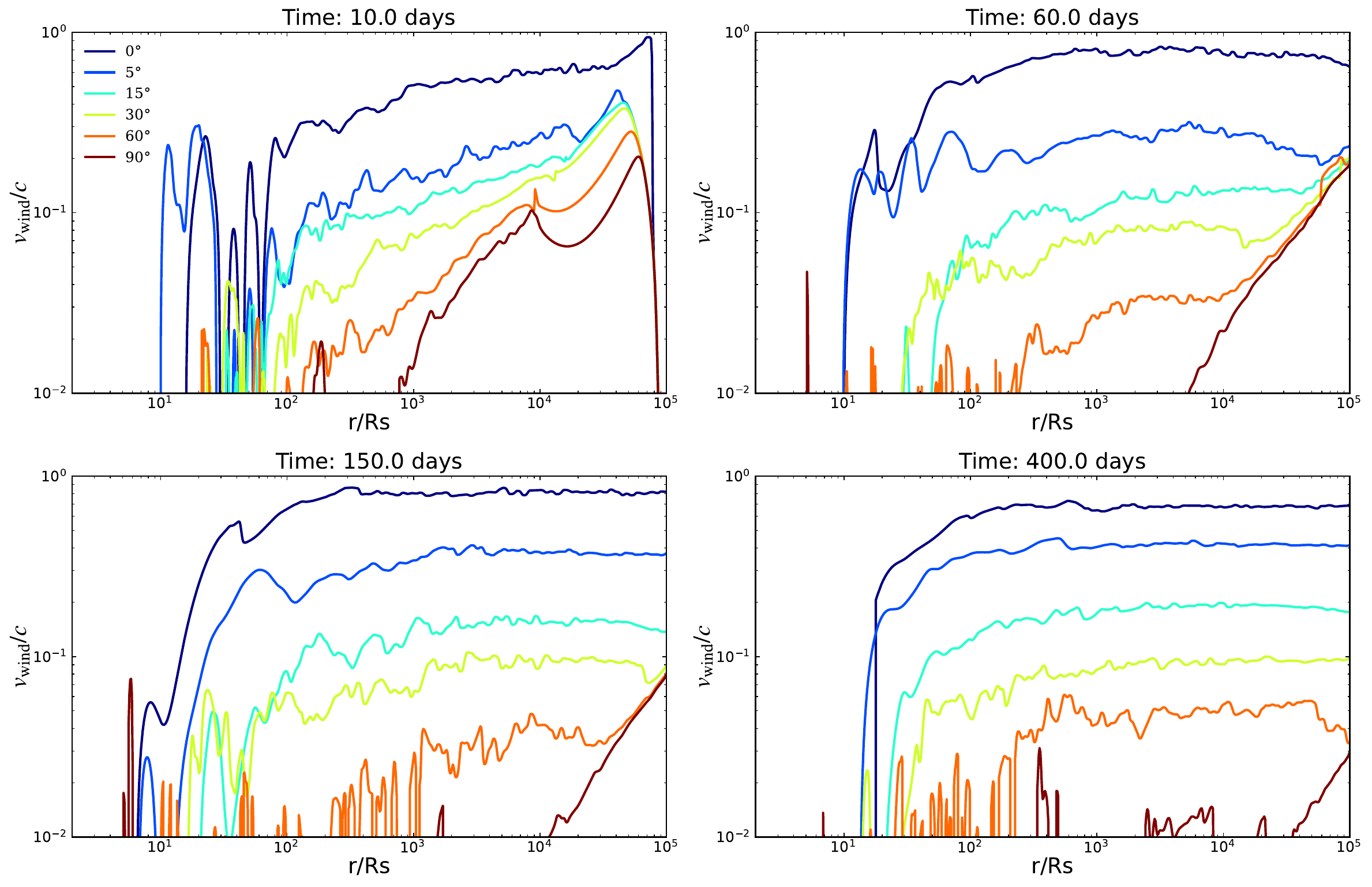}

\caption{
        Radial profile of radial velocity of the wind for model M300-9 at $t = 10.0\,\mathrm{\ days}$ (top-left panel),$t = 60.0\,\mathrm{\ days}$ (top-right panel), $t = 150.0\,\mathrm{\ days}$ (bottom-left panel) and $t = 400.0\,\mathrm{\ days}$ (bottom-right panel). In each panel we plot the velocity along six different viewing angles.
}
\label{fig:vtheta300-9}
\end{figure}

Fig.~\ref{fig:evol300-9} presents snapshots for the gas density and streamlines in the zoom-in area and zoom-out area for this system. A strong optically thick wind is driven by radiation pressure. We are particularly interested in the location of the scattering photosphere, where the electron scattering optical depth $\tau_{\rm es}$ is equal to unity. Initially, with time evolution, the photosphere expands with the outward movement of wind. Since the velocity of the wind increases with decreasing $\theta$ (see Fig. \ref{fig:vtheta300-9}), the vertical expansion of the photosphere is much faster than that in the horizontal direction. Therefore, initially, the photosphere has a vertically elongated morphology (see the top-third panel of Fig. \ref{fig:evol300-9}). At $t=60$ days, the photosphere expands to $\sim 70000$ $R_{\rm s}$ vertically, while at the mid-plane, the photosphere only reaches $\sim 35000R_{\rm s}$.  

Over time, the morphology of the photosphere gradually becomes horizontally elongated (see the lower-third panel of Fig. \ref{fig:evol300-9}). The photosphere close to the rotational axis gradually moves towards the black hole. This evolution is driven by the density decreasing with time. This makes the photosphere around the rotational axis move towards the center at late times. However, the wind mass flux around the mid-plane is still large even at late times, which can maintain a large photosphere. The morphology of the photosphere becomes horizontally elongated at late times. We find that the location of the photosphere at the rotational axis reaches its maximum value of $3.4\times10^{4}R_{\rm s}$ at $t= 21.8$ days. After that the photosphere at a viewing angle along rotational axis moves inward. At $t = 449$ days, the photosphere at $\theta = 0^\circ$ recedes to 2$R_{\rm s}$. Our results suggest that for Pop III star TDEs, the accretion disk may be heavily obscured during the early phases (t $\lesssim 449 \mathbf{\ days}$), even when observed along the rotational axis. 
In contrast, simulations by \cite{BU22} for a solar-type star TDE with a $10^6 M_\odot$ black hole have found that there is a critical angle $\theta_{\rm cr} \approx 10^\circ$. When the viewing angle satisfies $\theta_{\rm obs} < \theta_{\rm cr}$, the accretion system is always totally un-obscured and observable.

\begin{figure}[htbp]
\centering
\includegraphics[width=0.8\textwidth]{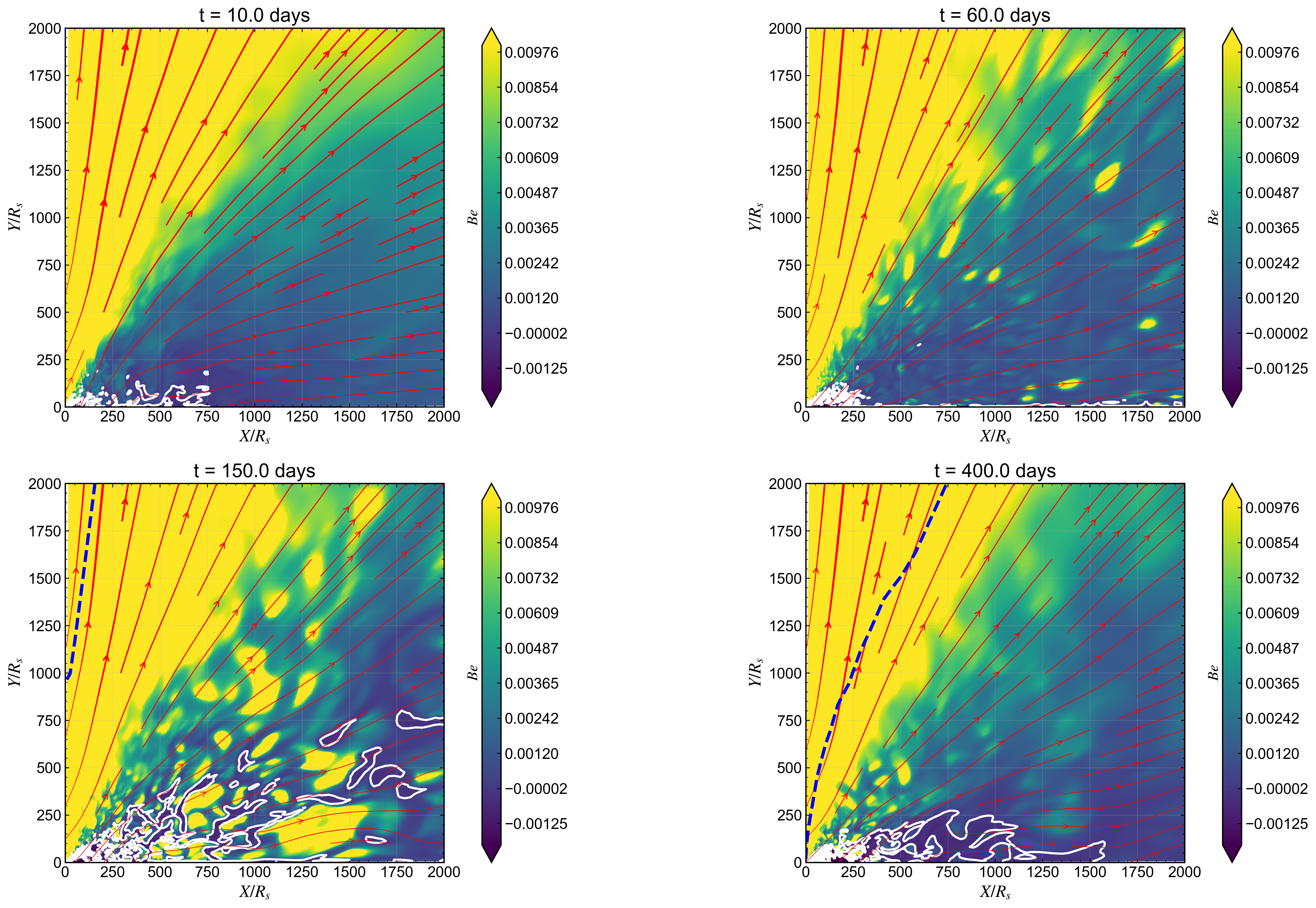}

\caption{
       Snapshots for the Bernoulli parameter (colour scale) with fluid velocity (streamlines) for model M300-9 at $t = 10.0\mathrm{\ days}$ (top-left panel), $t = 60.0\mathrm{\ days}$ (top-right panel), $t = 150.0\mathrm{\ days}$ (bottom-left panel) and $t = 400.0\mathrm{\ days}$ (bottom-right panel). The Bernoulli parameter is calculated in the code unit with $GM_{\rm BH} = R_{\rm s} = 1$. The white lines are the contour for $Be=0$. The black dashed lines in the two bottom figures depict the electron scattering photosphere.
}
\label{fig.be300-9}
\end{figure}

\begin{figure}[htbp]
\centering
\includegraphics[width=0.9\textwidth]{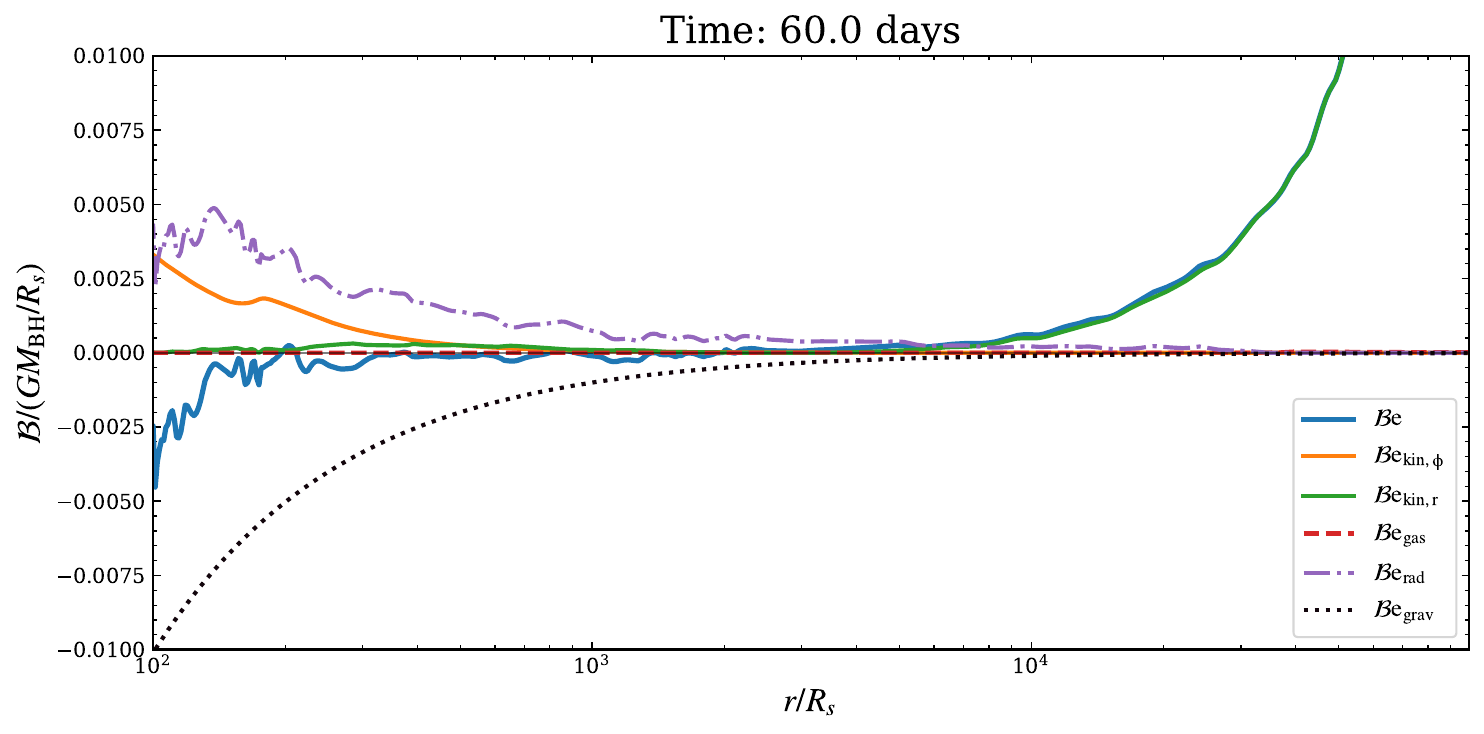}

\caption{
       Radial profiles of Bernoulli parameter  along the mid-plane at $t = \mathrm{60.0\ days}$ for model M300-9 . The blue solid line shows the Bernoulli parameter. The orange solid line, green solid line, red dashed line, purple dash-dotted line and black dotted line represent the rotation kinetic energy, the outflow kinetic energy, the gas enthalpy, the radiation energy enthalpy and the gravitational potential energy, respectively.
}
\label{fig.bp300-9}
\end{figure}

In Fig.~\ref{fig.be300-9}, we show the Bernoulli parameter in four snapshots ($t = $ 10, 60, 150, 400 days). The Bernoulli parameter is calculated in units with $G = M_{\rm BH} = R_{\rm s} = 1$. Three characteristic features can be seen from this figure. First, generally, most of the flow has positive Bernoulli parameter. Second, at a fixed radius $r$, the Bernoulli parameter generally increases with decreasing $\theta$. Third, at a fixed $\theta$, the Bernoulli parameter increases with increasing radius. All these features are similar to those found in \cite{BU22}. Now, we study how the Bernoulli parameter changes with radius for a fixed $\theta$. We take the Bernoulli parameter along the mid-plane as an example. In Fig.~\ref{fig.bp300-9}, we plot the radial profile of the Bernoulli parameter along the mid-plane at $t = 60$ day. In the region $2R_{\rm s} < r<6000 R_{\rm s}$, the dominant positive component of $Be$ is the radiation enthalpy. The kinetic energy due to $v_\phi$ dominates the kinetic energy due to $v_{\rm r}$ inside $500R_{\rm s}$. With the increase of radius, due to angular momentum conservation, $v_\phi$ decreases. Therefore, the kinetic energy component $1/2v_\phi^2$ decreases. However, due to the continuous acceleration of wind, the kinetic energy component $1/2v_{\rm r}^2$ gradually increases with radius. Outside $6000R_{\rm s}$, the component $1/2v_{\rm r}^2$ becomes completely dominating. In the region $r<6000R_{\rm s}$, with the increase of radius, the decrease of radiation enthalpy mainly compensates the increase of gravitational energy. The Bernoulli parameter increases with radius and it transits from negative to positive values at $\sim 2000 R_{\rm s}$. The Bernoulli parameter is not conserved due to the fact that the flow is neither laminar nor inviscid.

We now quantitatively characterize the wind properties and their temporal evolution. In our simulation, the wind is accelerated by radiation pressure. Previous studies have found that the wind could be accelerated to $\geq 0.6c$ near the rotational axis (e.g., \cite{Yoshioka2022}) by radiation pressure when the accretion rate exceeds $\sim 100\dot{M}_{\rm Edd}$. Fig. \ref{fig:vtheta300-9} shows the time evolution of the radial profiles of the radial velocity of the wind. In this model, gas is injected at a radius of $R_{\rm C}=34.68R_{\rm s}$. In the region around $R_{\rm C}$, the properties of the wind may be quite affected. We focus on the properties of the wind in the region $r > 100 R_{\rm s}$. This region is far away from the gas injection region and initially there is no wind at all. Generally, the velocity of the wind decreases with increasing $\theta$ as found in previous works (e.g., \cite{Yang2023}, \cite{Yuan2015}, \cite{BU23}, and \cite{Dai2018}). In our simulation, the acceleration caused by radiation is more pronounced than that in previous works due to the extremely high accretion rate. Maximum wind velocities can be as high as $\sim 0.9c$ near the polar axis. At the mid-plane, the velocity of the wind can be $\sim 0.2c$ at an early time and decreases with time. The velocity of the wind is significantly higher than the local escaping velocity. Therefore, when wind moves outward, it can hardly be decelerated by the black hole gravity. We can see that in the region $r> 100R_{\rm s}$, the wind velocity even increases with radius at $t=10$ days. At $t\geq 60$ days, the wind velocity is almost a constant of radius in the region $r> 100R_{\rm s}$. As the speed of wind can be as high as $0.9 c$, special relativity effects may become important to affect wind properties. Therefore, future studies incorporating special relativistic effects are necessary.

\begin{figure}[htbp]
\centering
\includegraphics[width=0.8\textwidth]{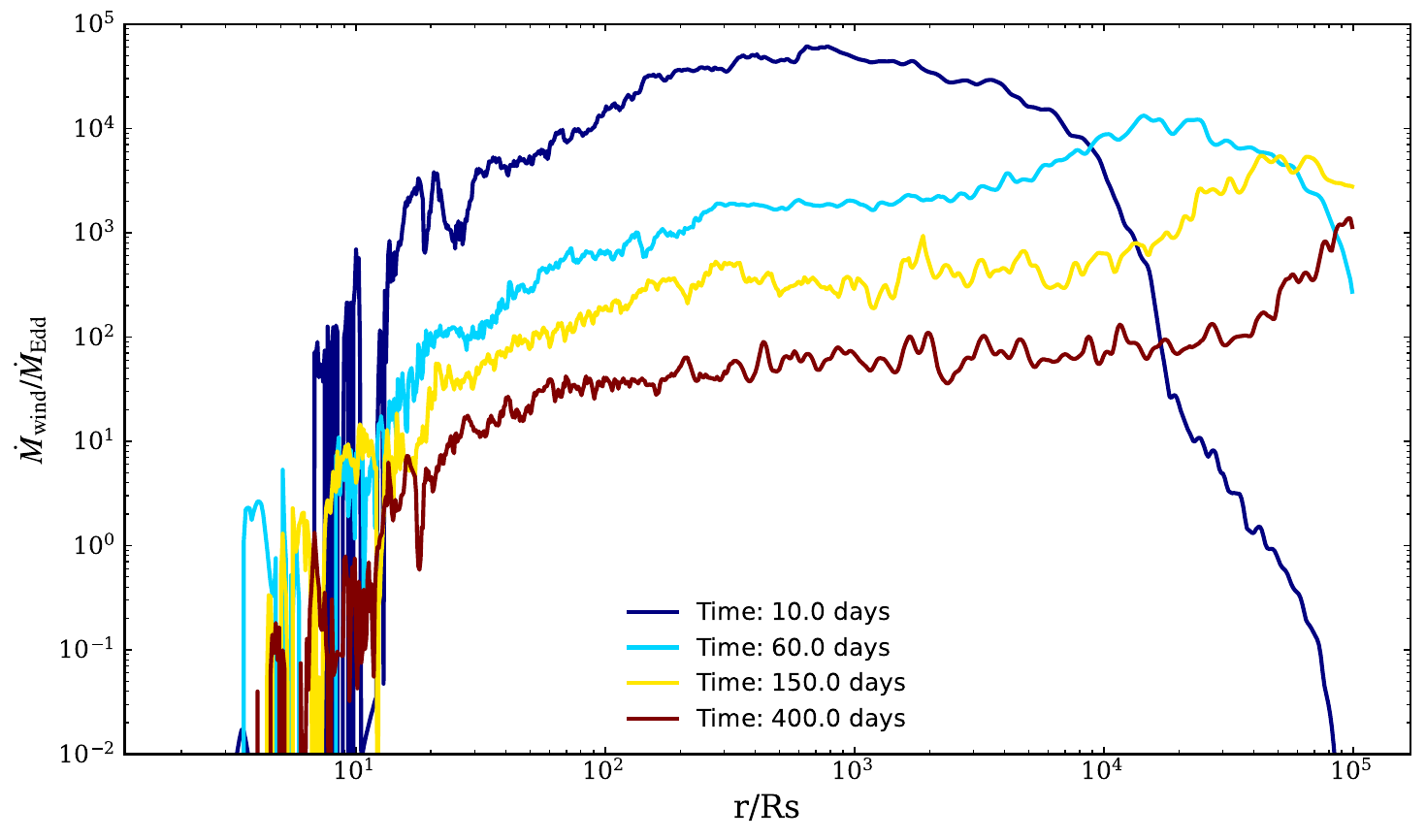}

\caption{
        Radial profiles of mass flux of wind for model M300-9. The blue, cyan, yellow and red lines denote the profile at $t = 10.0\,\mathrm{\ days}$, $t = 60.0\,\mathrm{\ dasy}$, $t = 150.0\,\mathrm{\ days}$ and $t = 400.0\,\mathrm{\ days}$ respectively.
}
\label{wp300-9}
\end{figure}

In Fig.~\ref{wp300-9}, we show the radial profiles of the wind mass flux at different snapshots. In the region $\sim10R_{\rm s}$, due to the strong gravity, the wind is weak and fluctuating.  Previous simulations of both sub-Eddington hot accretion flow and super-Eddington accretion flow have all found that the wind is extremely weak inside $\sim10R_{\rm s}$ (\cite{2012ApJ...761..130Y}; \cite{Curd&Narayan2019}). In the region far away from $R_{\rm C}=34.68R_{\rm s}$, the profiles are quite smooth. This figure shows two important features of the wind mass flux. First, generally the mass flux of the wind decreases with time. The reason is that almost a fixed fraction of the injected gas is accreted (see Fig. \ref{dot300-9}). The other fraction of the injected gas is launched as wind. The injection rate declines with time. Therefore, the wind mass flux decreases with time. Second, at a given snapshot, generally, the wind mass flux first increases and then decreases with radius. The reason is as follows: We take the snapshot at $t=10$ days as an example. Initially, there is no wind in the computational domain at all. With the injection of gas and the formation of the accretion flow, wind is driven by radiation pressure. The gas injection rate is highest at $t=0$. Also, the wind is strongest initially. At $t=10$ day, the initially generated wind has arrived at $\sim 3000 R_{\rm s}$. Outside $\sim 3000 R_{\rm s}$, due to the limited evolution time, the wind has no sufficient time to arrive there. Therefore, outside $\sim 3000 R_{\rm s}$, the mass flux of the wind quickly decays to very small value with the increases of radius. Inside $3000R_{\rm s}$, the mass flux of the wind decreases with the decreasing radius. This is because the wind is defined as gas with both positive $v_{\rm r}$ and Bernoulli parameter. There is gas with positive $v_{\rm r}$ and negative Bernoulli parameter. This portion of gas is not classified as wind. However, as mentioned above (see also Fig. \ref{fig.bp300-9}), when gas moves outwards, the Bernoulli parameter can increase and transit from negative to positive value. Therefore, more and more gas is recorded as wind with the increase of radius. Besides that, when considering a snapshot, wind observed at larger radii was actually launched at an earlier time when the accretion rate was higher. Since the wind mass flux is proportional to the accretion rate at the launch time, the wind mass flux measured at a larger radius can appear higher in such instantaneous profiles.  

\begin{figure}[htbp]
\centering
\includegraphics[width=0.8\textwidth]{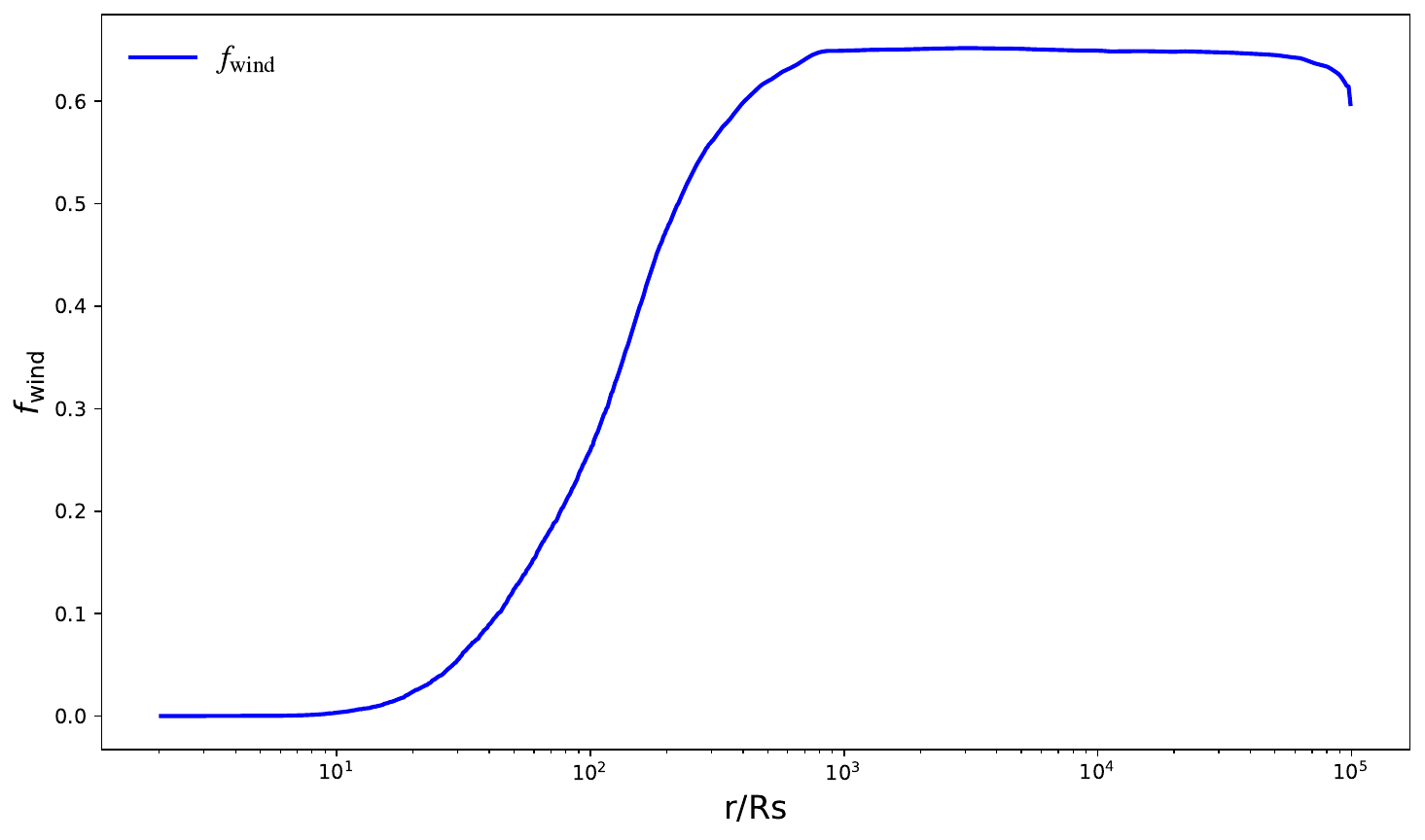}

\caption{
    The radial profile of the ratio of the time-integrated mass taken away by the wind to the mass of the injected gas over 499.89 days for model M300-9.
}
\label{fig.f300-9}
\end{figure}

An important question is that how much of the fallback debris is taken away by the wind. In order to answer this question, in Fig.~\ref{fig.f300-9}, we present the radial distribution of the ratio of the mass taken away by the wind to the mass of the injected gas over 499.89 days. This ratio is calculated as
\begin{equation}
f_{\mathrm{wind}}(r) = \frac{M_{\mathrm{wind}}(r)}{M_{\mathrm{fallback}}} = \frac{\int_{0}^{499.89\,\mathrm{d}} \dot{M}_{\mathrm{wind}}(r)  dt}{\int_{0}^{499.89\,\mathrm{d}} \dot{M}_{\mathrm{fb}}  dt}
\label{eq:wind_fraction}
\end{equation}

The value of $f_{\rm wind}$ increases with radius within $r \lesssim 800 R_{\rm s}$ and reaches a maximum value of $\sim 0.65$. The increase of $f_{\rm wind}$ with radius is also caused by the transition of Bernoulli parameter. The value of $f_{\rm wind}$ is almost a constant from $800 R_{\rm s}$ to $\sim 6\times10^{4} R_{\rm s}$. This indicates that in an integrated sense over time, the transition of the Bernoulli parameter of the outward moving gas from negative to positive value finishes roughly inside $800R_{\rm s}$. We note that for a specific snapshot and  angle $\theta$, the actual transition location can deviate from 800 $R_{\rm s}$ (for example, it can be seen from Fig.~\ref{fig.bp300-9}, such transition happens at a much larger scale). The value of $f_{\rm wind}$ declines slightly with radius outside $6\times10^{4} R_{\rm s}$. The reason is that it takes time for the wind to move from small to large radii. Therefore, in a time period from the beginning of the simulation, there should be no wind at large radii. Therefore, when we calculate $f_{\rm wind}$ by Equation \ref{eq:wind_fraction}, there is a decrease of $f_{\rm wind}$ with radius outside some specific radius, which is $6\times10^{4} R_{\rm s}$. Roughly  $\sim 65\%$ of the fallback debris is taken away by gas. This is consistent with the above result that roughly $35\%$ of the fallback debris is accreted by the black hole.

\begin{figure}[htbp]
\centering
\includegraphics[width=0.9\textwidth]{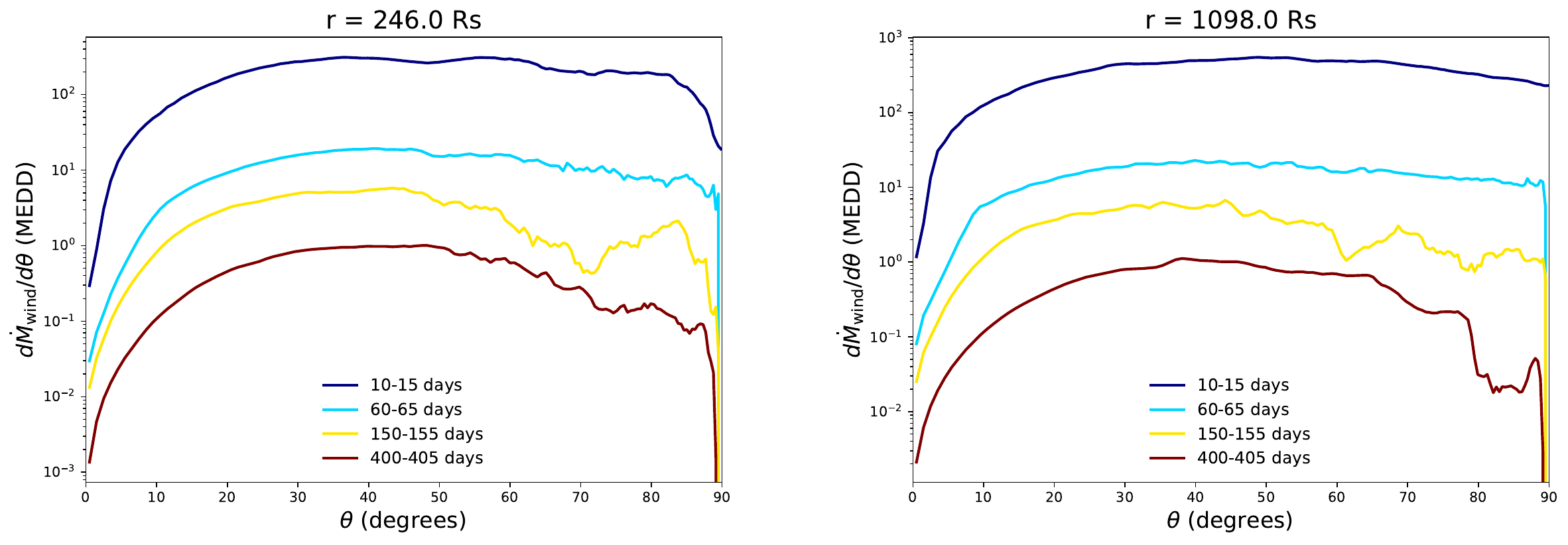}

\caption{
    The angular($\theta$) distribution of mass flux of wind in unit of Eddington accretion rate for model M300-9 at $r = 246.0\mathrm{R_s}$ (left panel) and $r = 1098.0\mathrm{R_s}$ (right panel). In order to eliminate the fluctuation, we do time average to the wind mass flux. The blue, cyan, yellow and red lines represent average period of 10-15, 60-65, 150-155 and 400-405 days respectively.
}
\label{fig.wt300-9}
\end{figure}

When calculating the radio emission generated by the interaction between the wind and the CNM or interstellar clouds, the angular distribution of the wind is an important ingredient. Therefore, it is important to study the angular distribution of the mass flux and kinetic power of the wind. In Fig.~\ref{fig.wt300-9} we show the angular distribution of wind mass flux. To eliminate fluctuations, we time-averaged the results over 5-day periods around different time snapshots and plot the results at two locations: $246 R_{\rm s}$ and $1098 R_{\rm s}$. The wind mass flux is significantly lower near the polar axis, which is attributable to the lower gas density in that region. It increases steeply between $0^\circ$ and $10^\circ$ and then more gradually up to $30^\circ$. With the increase of $\theta$, the density increases while the velocity decreases. Thus the mass flux is roughly constant in the region of $30^\circ-80^\circ$. Similar angular distributions appear in recent numerical simulations of super-Eddington accretion flows (e.g., \cite{Yang2023}, \cite{BU23}).

\begin{figure}[htbp]
\centering  
\includegraphics[width=0.9\textwidth]{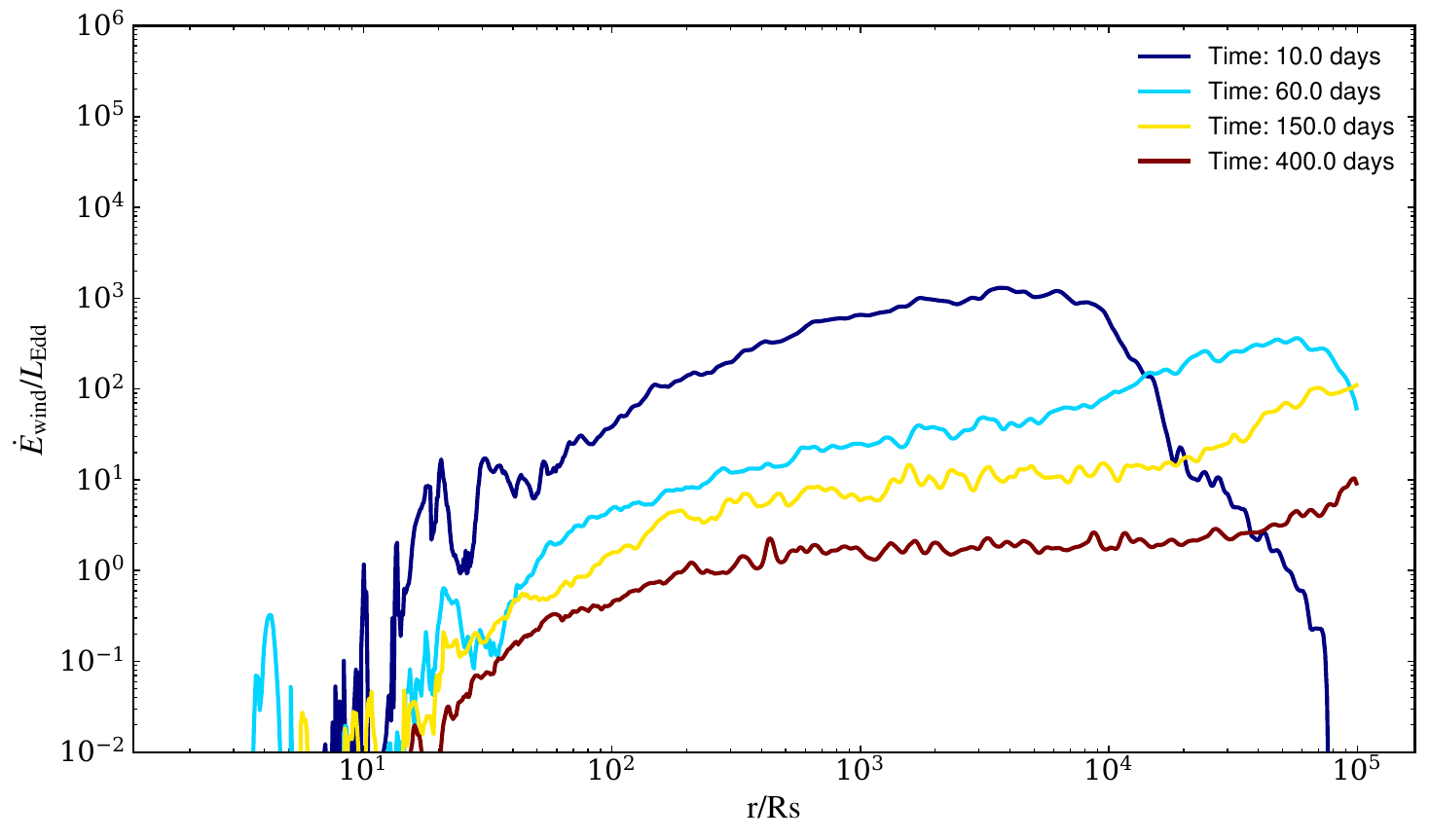}
\caption{
       Radial profiles of kinetic power of wind for model M300-9. The blue, cyan, yellow and red lines denote the profile at $t = 10.0\,\mathrm{\ days}$, $t = 60.0\,\mathrm{\ days}$, $t = 150.0\,\mathrm{\ days}$ and $t = 400.0\,\mathrm{\ days}$ respectively.
       }
\label{fig.ep300-9}
\end{figure}

The wind can interact with the black hole surrounding environment and produce echoes at different wavelength bands. For example, recent works show that in TDEs in the local universe, the radio emission of non-jetted TDEs may be due to the interaction of the wind and the black hole surrounding medium (e.g., \cite{1998ApJ...499..810C}; \cite{2013ApJ...772...78B}; \cite{2021MNRAS.507.4196M}; \cite{Mou2022}; \cite{2023MNRAS.521.4180B}; \cite{2025ApJ...979..109Z}). In the radio emission production process, the wind power is an important ingredient. Therefore, it is very interesting to study wind power. In Fig.~\ref{fig.ep300-9} we plot the radial distribution of the kinetic power carried by the wind at four time snapshots. Similarly to the behaviour of the wind mass flux, generally the kinetic power of the wind also decreases with time. The maximum kinetic power can be $10^3L_{\rm Edd} \sim 10^{47} {\rm erg\cdot s^{-1}}$. We note that for a solar-type star TDE with a $10^6M_\odot$ black hole, we find that the kinetic power of the wind is $\sim 10^{44} {\rm erg\cdot s^{-1}}$. Such a powerful wind should have significant effects on the black hole surroundings.

\begin{figure}[htbp]
\centering 
\includegraphics[width=0.9\textwidth]{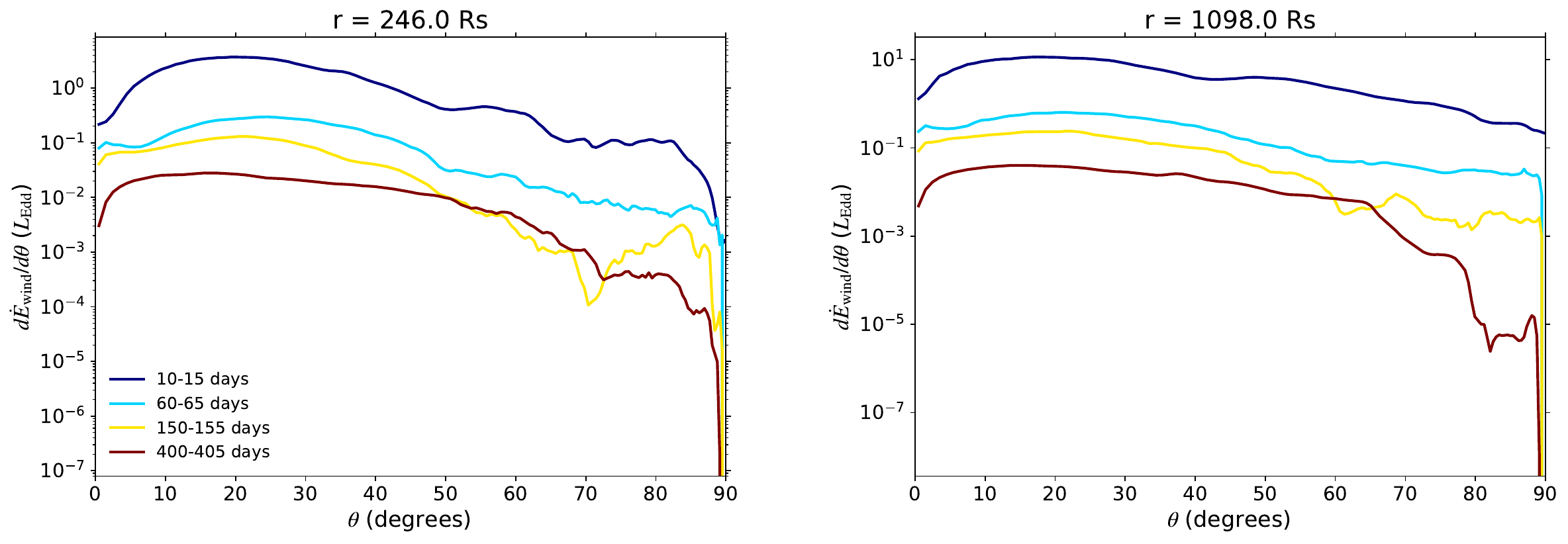}

\caption{
    The angular($\theta$) distribution of kinetic power of wind in unit of $L_{\rm Edd}$ for M300-9 at $r = 246.0\mathrm{R_{\rm s}}$ (left panel) and $r = 1098.0\mathrm{R_{\rm s}}$ (right panel). In order to eliminate the fluctuation, we do time average to the kinetic power. The blue, cyan, yellow and red lines represent average period of 10-15, 60-65, 150-155 and 400-405 days respectively.
}
\label{fig.et300-9}
\end{figure}

The angular distribution of the wind kinetic power is important to determine the interaction efficiency between the wind and black hole surroundings. The interaction is more efficient for the wind with a higher opening angle. If the kinetic power of the wind is confined to a very small angle, the interaction will not be sufficient even for  a wind with significantly high power. We show the angular distribution of the kinetic power carried by the wind at $r = 246 R_{\rm s}$ and $r = 1098 R_{\rm s}$ in Fig.~\ref{fig.et300-9}. We averaged the data over 5-day intervals around four time snapshots to eliminate fluctuations. Although wind velocities are extremely high near the polar axis ($\theta \approx 0^\circ$), the low mass flux in this region results in relatively low kinetic power. As the mass flux increases (as shown in Fig.~\ref{fig.wt300-9}) with $\theta$, the kinetic power rises, peaking around $\theta \sim 20^\circ$. Beyond this angle, despite a near-constant mass flux extending towards the disk plane, the kinetic power declines rapidly due to decreasing wind velocities. 

\begin{figure}[htbp]
\centering
\includegraphics[width=0.9\textwidth]{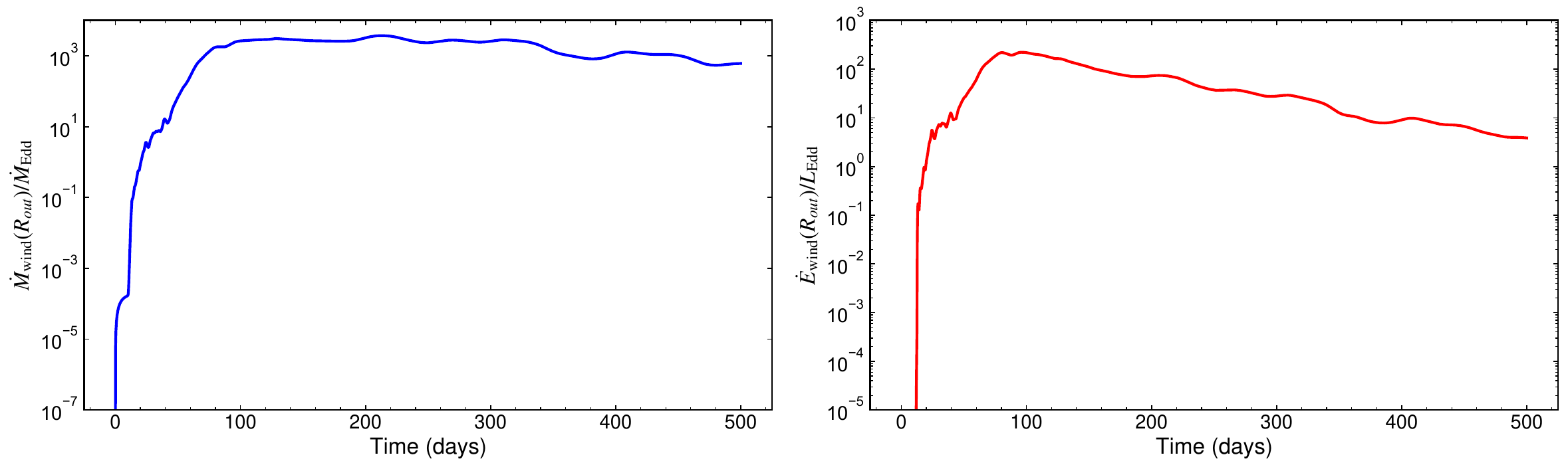}

\caption{
The temporal profile of the mass flux (left panel) and kinetic power (right panel) of wind for model M300-9 measured at the simulation
outer boundary $10^5R_{\rm }$.}
\label{fig.es300-9}
\end{figure}

Finally, in Fig.~\ref{fig.es300-9} we present the time evolution of the wind mass flux and the kinetic power at the outer radial boundary $10^5R_{\rm s}$. The information is important to study the interaction of the wind and the black hole ambient medium. The wind head arrives at the outer boundary roughly at $t \sim 15\mathrm{\ days}$. After that, both the wind mass flux and kinetic power increase rapidly over time and reach their peak values at $t \approx 100$ days. The maximum mass flux reaches $\sim 2.64\times$ $10^3 \dot{M}_{\rm Edd}$, and persists at this level during long-term evolution. After $t \approx 320\mathrm{\ days}$ the mass flux decreases due to the decrease in injected mass, reaching $6.08\times10^2\dot{M}_{\rm Edd}$ at $t =499.89\mathrm{\ days}$.
Meanwhile, the kinetic power reaches a maximum value of $\sim 218 L_{\rm Edd}$.  After that time, the kinetic power of the wind gradually decreases with time, reaching a value of 3.86$L_{\rm Edd}$ at $t =500\mathrm{\ days}$. The maximum values of $\dot{M}_{\rm wind}(R_{\rm out})$ and $\dot{E}_{\rm wind}(R_{\rm out})$ are smaller than the peak values shown in Fig.~\ref{wp300-9} and Fig.~\ref{fig.ep300-9}. This is because the radial velocity increases with increasing radii at an early time for a large fraction of the wind, as shown in Fig.~\ref{fig:vtheta300-9}. The difference in velocities of the front and rear ends of the wind causes the winds to expand while traveling outside, making the peak of mass flux decrease. To explain this clearly, we could define $\dot{M}_0$ as the peak mass flux of the wind at time $\rm t = t_0$ and consider that the winds near its location have a width noted as $\Delta_0$. We could simply note the velocities of the front and rear ends of the wind as $v = v_0 \pm \delta_v$, where $v_0$ is the mean velocity of this part of winds. At time $t = t_0 + dt$, the width of this part of the wind becomes $\Delta_1 = \Delta_0 + 2\delta_vdt$ so that the mass flux becomes $\dot{M}_1 \simeq \dot{M}_0\cdot\Delta_0/\Delta_1$ that decreases.

\subsection{Wind in Pop III star TDEs with $M_\star = 300 M_\odot$ and metallicity $Z= 10^{-5} Z_\odot$ (Model m300-5)}
\begin{figure}[htbp]
\centering  
\includegraphics[width=1.0\textwidth]{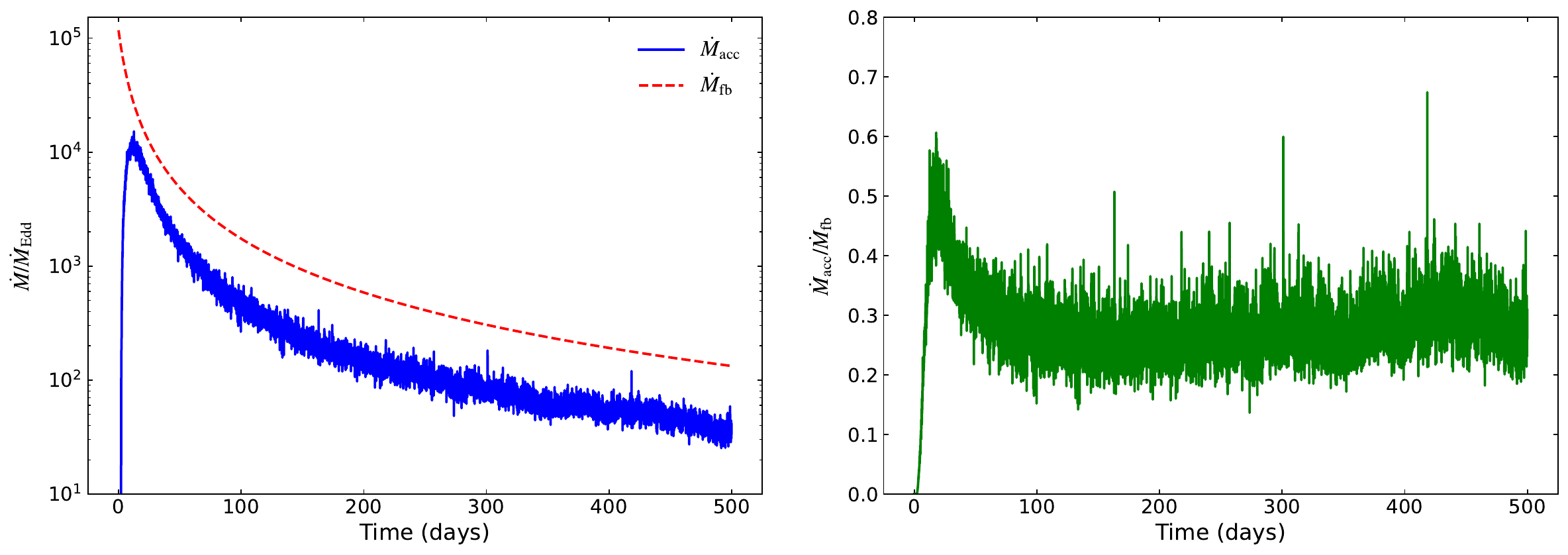}

\caption{Accretion rate for model M300-5. Left panel: time evolution of the black hole accretion rate (blue line) and stellar debris fallback rate 
(red dotted line) in Eddington units. Right panel: time evolution of black hole accretion in the unit of the stellar debris fallback rate.
}
\label{fig.mdfb300-5}
\end{figure}

 In this section, we study the model with a metallicity of $10^{-5}Z_\odot$. We highlight characteristic outcomes of model M300-5, directly comparing them with the results of model M300-9 presented in the previous section.
The fallback time scale, as calculated by Eq.~\ref{eq11}, is $\sim$8.7 days for model M300-5 and $\sim$3.5 days for model M300-9. The peak fallback rates (or gas injection rates) are $1.17\times 10^{5}\dot{M}_{\rm Edd}$ for model M300-5 and  $2.95\times 10^{5}\dot{M}_{\rm Edd}$ for model M300-9. A calculation using Eq.\ref{eq6} shows that the mass injection rate in model M300-5 exceeds that in model M300-9 at $t = 8.3\mathrm{\ days}$. However, if we integrate the injection rate over 500 days, we will find that the overall injected mass of model M300-9 is still higher. During 500 days, $\sim 89.6M_{\odot}$ is injected for model M300-9 and $\sim 86.7M_{\odot}$ is injected for model M300-5. Due to the extremely high inject rate, $\sim 51.1M_{\odot} $ is injected into the system in the early 9 days for model M300-9 and $\sim 32.8M_{\odot} $ is injected for model M300-5. In model M300-9, $\sim 1/3$ of the debris mass is injected into the simulation domain in the first 9 days, triggering an extremely high accretion rate and driving powerful winds. However, the evolution of model M300-5 is quite slower.

In Fig.~\ref{fig.mdfb300-5}, we show the time evolution of stellar debris fallback and accretion rate (left panel) and the ratio of accretion rate to the debris fallback rate (right panel) for this model. Similarly to Fig.~\ref{dot300-9}, the accretion rate of model M300-5 undergoes an initial increase and peaks at $\sim13\mathrm{\ days}$ with a peak value of $1.53\times10^4\dot{M}_{\rm Edd}$ followed by a gradual decay. The peak time for this model also coincides with the dynamic timescale ($\sim 10\ \rm days$), which is shorter than the viscous timescale ($\sim18 \ \rm days$). The underlying reason is the same as that introduced previously. Since the initial debris injection rate calculated from Eq.~\ref{eq12} is lower than that in model M300-9, the peak value of the accretion rate is lower. The ratio of accretion rate to injection rate tends to stabilize at a value $\sim0.25$. Comparing Fig.\ref{dot300-9} and Fig.\ref{fig.mdfb300-5}, we could find that during about the early 200 days the accretion rate in model M300-9 is higher; while after 200 days the accretion rate in model M300-5 becomes higher.

The disk in model M300-5 grew larger than that in model M300-9. Its outer boundary increases throughout the 500-day simulation and  reaches $\sim 1000\ R_{\rm S}$. This greater expansion was due to the larger injection radius, which provided the debris with a higher initial specific angular momentum.

\begin{figure}[htbp]
\centering  
\includegraphics[width=1.0\textwidth]{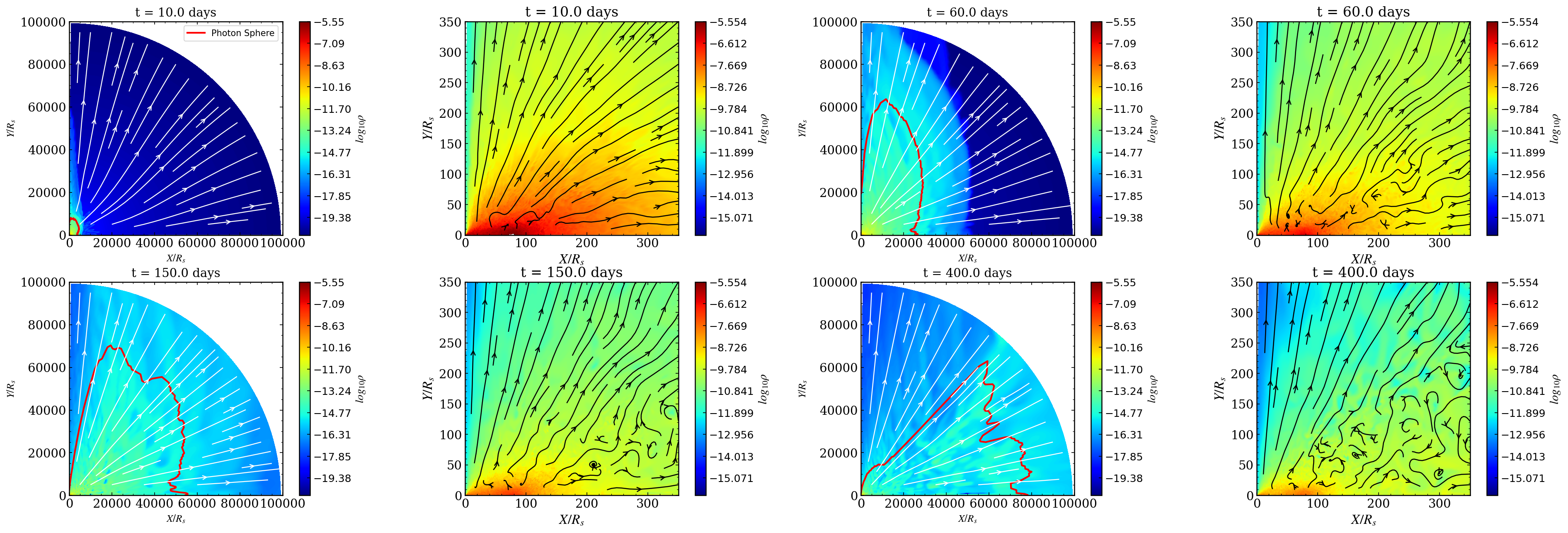}

\caption{
       Snapshots for gas density (colour scale) for model M300-5 with fluid velocity (streamlines). We illustrate the system state at 
    $t = 10.0\,\mathrm{days}$ (top-first and top-second panels), $t = 60.0\,\mathrm{days}$ (top-third and top-fourth panels), $t = 150.0\,\mathrm{days}$ (bottom-first and bottom-second panels), and $t = 400.0\,\mathrm{days}$ (bottom-third and bottom-fourth panels). For each time snapshot, we present two panels to show the properties of a zoom-out large domain and a zoom-in smaller domain of the system. The red lines in the top-first, top-third, bottom-first and bottom-third panels are the electron scattering photosphere.}
\label{fig.evol300-5}
\end{figure}

Fig.~\ref{fig.evol300-5} shows the snapshots of the gas density in the zoom-in and zoom-out areas with a scattering photosphere and streamlines. There is also an optically thick wind driven by radiation pressure. The wind causes the photosphere to expand with time. The radial profiles of radial velocity in Fig.~\ref{fig:vtheta300-5} are similar to those in Fig.~\ref{fig:vtheta300-9}. The velocity of the wind increases with decreasing $\theta$, causing the photosphere to be elongated vertically at an early time. At $t=60\mathrm{\ days}$ the photosphere reaches $\sim 64000R_{\rm s}$ vertically and just reaches $\sim25000R_{\rm s}$ at the mid-plane. 
The morphology of the photosphere transitions from a vertically elongated shape to a horizontally elongated shape at late times for the same reason we mentioned above. At the end time of the simulation $t = 500$ days, the debris fallback rate in model M300-5 is higher than that in model M300-9. As a result, the location of the photosphere along the rotational axis is located at $\sim400R_{\rm s}$ at $t=500\mathrm{\ days}$. The entire accretion flow is enveloped by the wind. The X-rays generated very close to the black hole can not leak out even at $t=500\mathrm{\ day}$. As a note that in model M300-9, at $t=449$ days, the photosphere at the rotational axis recedes to 2$R_{\rm s}$.

\begin{figure}[htbp]
\centering  
\includegraphics[width=0.9\textwidth]{300-5vprofile_comparison.pdf}

\caption{
        Radial profile of radial velocity of the wind for model M300-5 at $t = 10.0\,\mathrm{\ days}$ (top-left panel),$t = 60.0\,\mathrm{\ days}$ (top-right panel), $t = 150.0\,\mathrm{\ days}$ (bottom-left panel) and $t = 400.0\,\mathrm{\ days}$ (bottom-right panel). In each panel we plot the velocity along six different viewing angles.}
\label{fig:vtheta300-5}
\end{figure}

\begin{figure}[htbp]
\centering  
\includegraphics[width=0.8\textwidth]{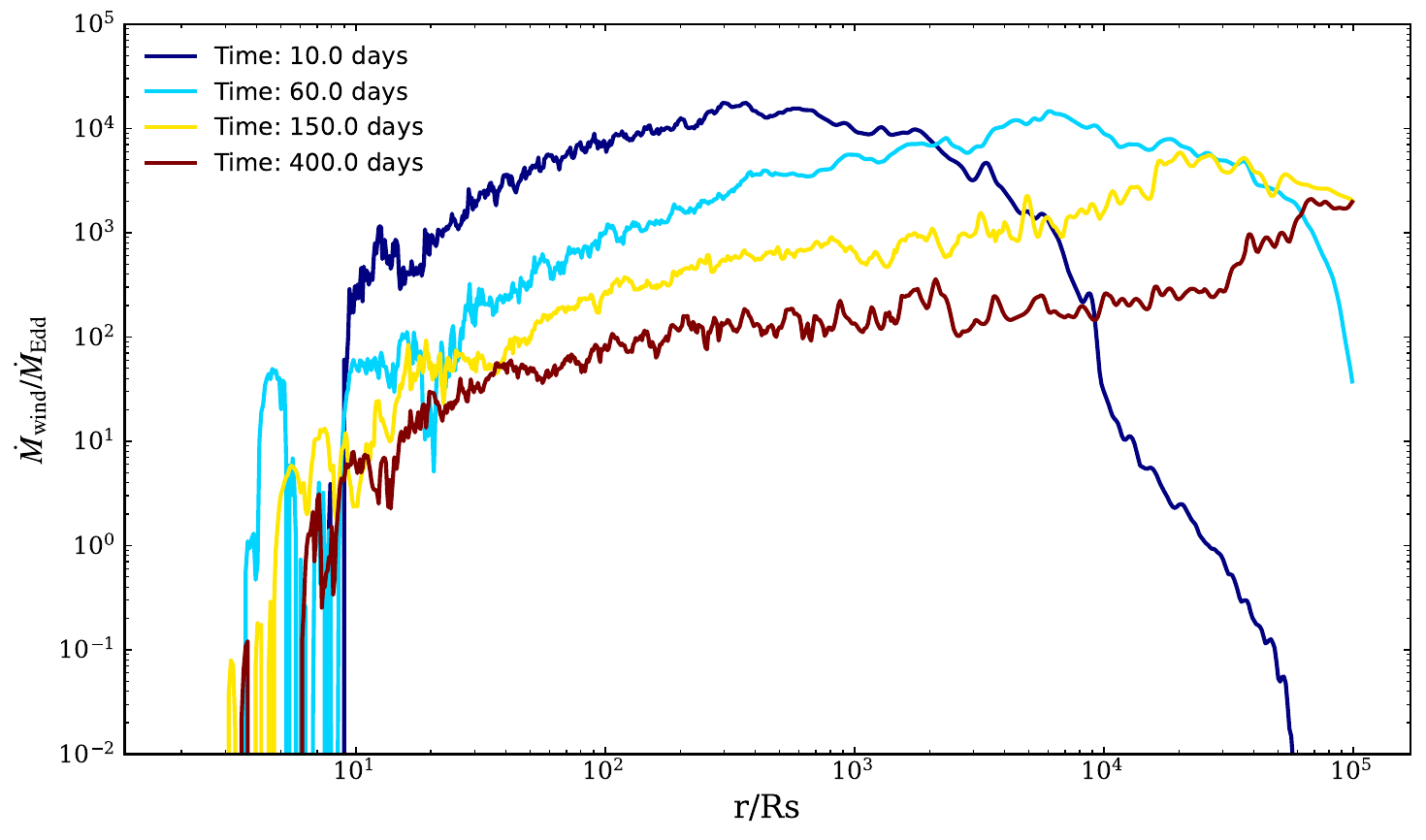}

\caption{
        Radial profiles of mass flux of wind for model M300-5. The blue, cyan, yellow and red lines denote the profile at $t = 10.0\,\mathrm{\ days}$, $t = 60.0\,\mathrm{\ dasy}$, $t = 150.0\,\mathrm{\ days}$ and $t = 400.0\,\mathrm{\ days}$ respectively.
}
\label{wp300-5}
\end{figure}

\begin{figure}[htbp]
\centering  
\includegraphics[width=0.9\textwidth]{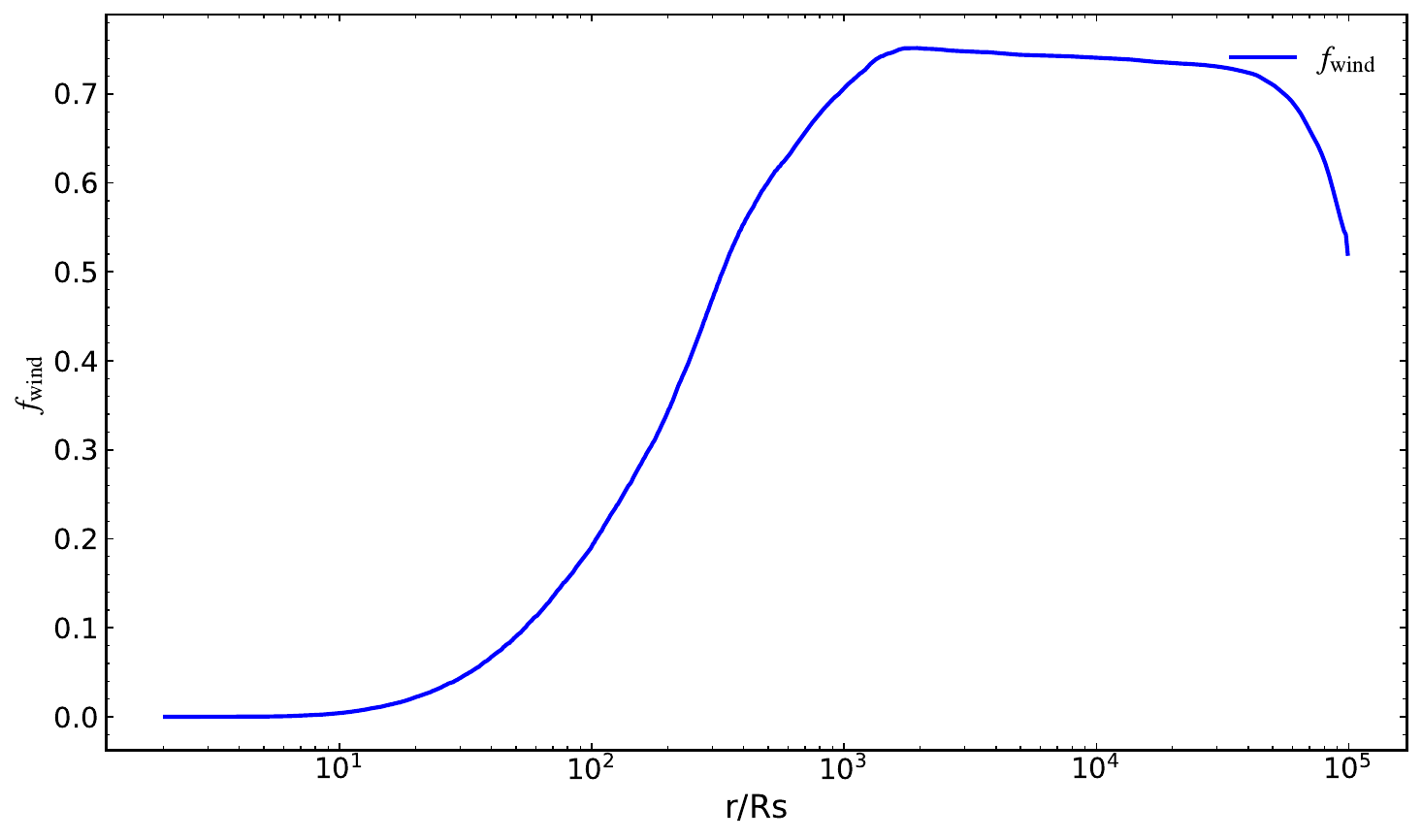}

\caption{
    The radial profile of the ratio of the time-integrated mass taken away by the wind to the mass of the injected gas over 499.89 days for model M300-5.
}
\label{fig.f300-5}
\end{figure}

\begin{figure}[htbp]
\centering  
\includegraphics[width=0.9\textwidth]{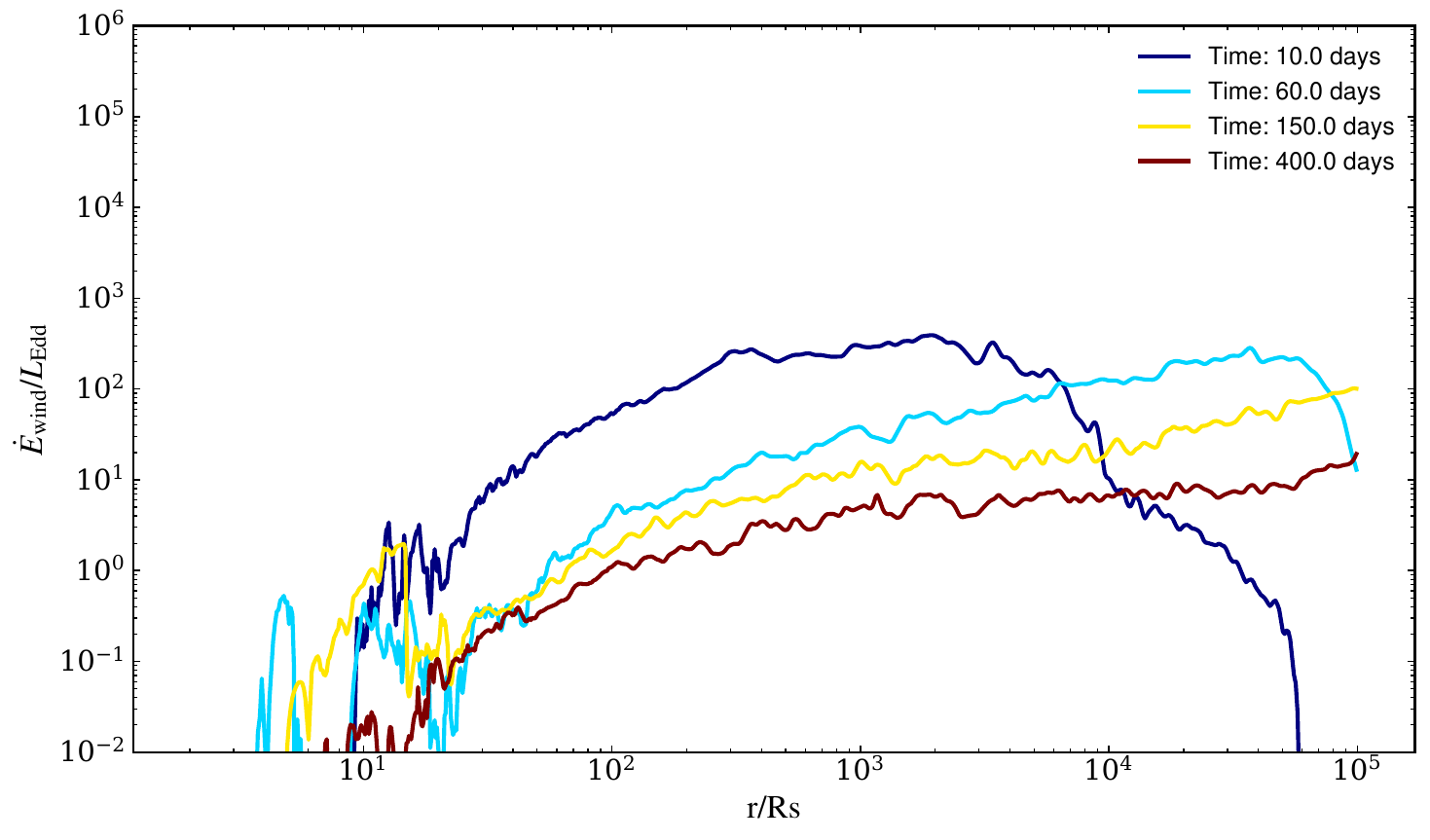}

\caption{
       Radial profiles of kinetic power of wind for model M300-5. The blue, cyan, yellow and red lines denote the profile at $t = 10.0\,\mathrm{\ days}$, $t = 60.0\,\mathrm{\ days}$, $t = 150.0\,\mathrm{\ days}$ and $t = 400.0\,\mathrm{\ days}$ respectively.
       }
\label{fig.ep300-5}
\end{figure}

\begin{figure}[htbp]
\centering 
\includegraphics[width=0.9\textwidth]{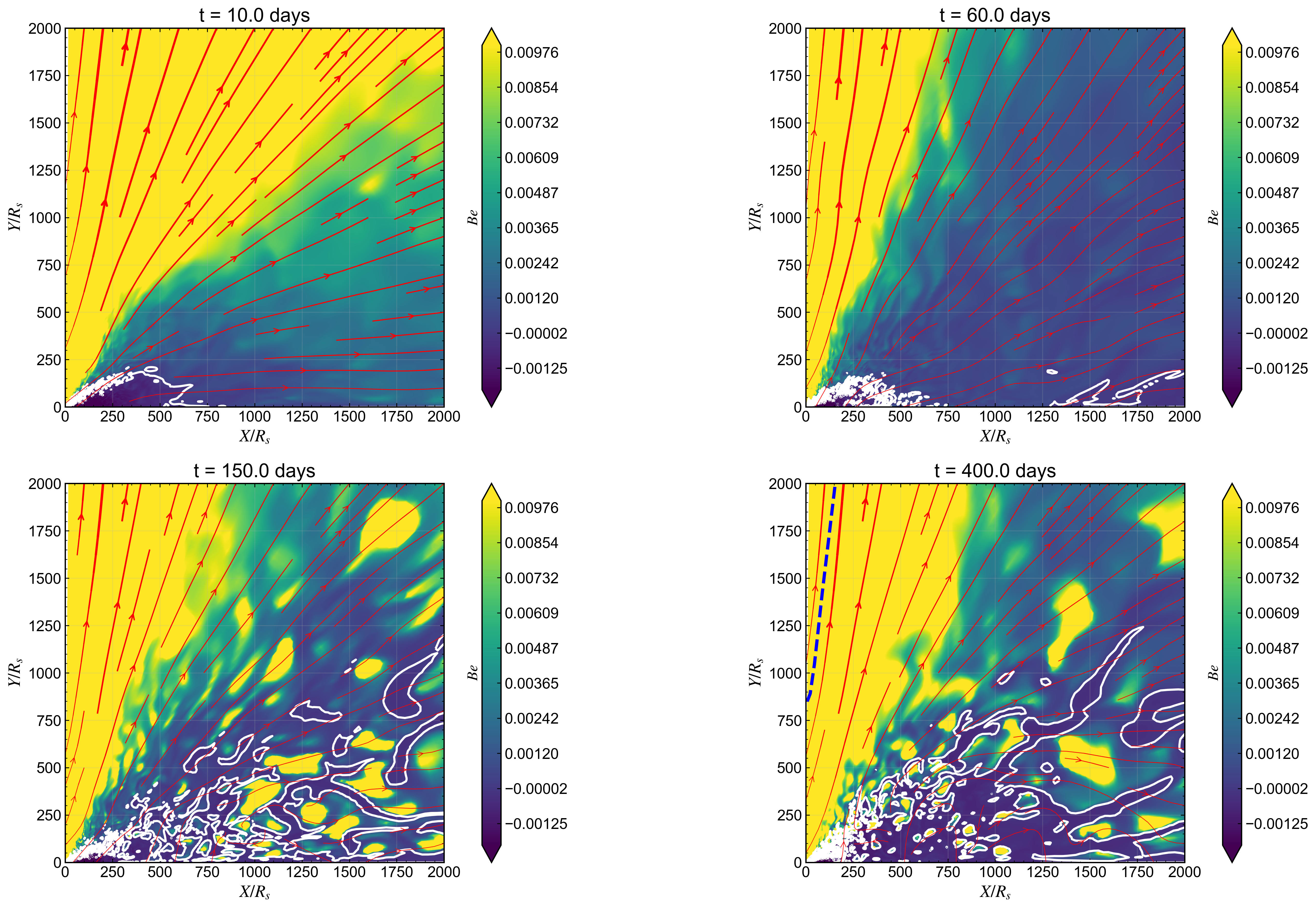}

\caption{
       Snapshots for the Bernoulli parameter (colour scale) with fluid velocity (streamlines) for model M300-5 at $t = 10.0\mathrm{\ days}$ (top-left panel), $t = 60.0\mathrm{\ days}$ (top-right panel), $t = 150.0\mathrm{\ days}$ (bottom-left panel) and $t = 400.0\mathrm{\ days}$ (bottom-right panel). The Bernoulli parameter is calculated in the code unit with $GM_{\rm BH} = R_{\rm s} = 1$. The white lines are the contour for $Be=0$. The black dashed lines in the bottom-right figure depict the electron scattering photosphere.
}
\label{fig.be300-5}
\end{figure}

In Fig.~\ref{wp300-5}, we plot the radial profiles of wind mass flux at different snapshots for model M300-5. In this model, the gas is injected around $R_{\rm C} = 79.44R_{\rm s}$, inside which the accretion flow is formed and wind can be generated. In the region $\sim10R_{\rm s}$, due to the strong gravity, the wind is weak. The radial profiles and their time evolution patterns are similar to those found in model M300-9.  The mass flux of the wind can reach a peak value $\sim 1.5\times 10^4\dot{M}_{\rm Edd}$ at $\sim 300R_{\rm s}$ at $t = 10.0\mathrm{\ days}$. It is slightly lower than the peak wind mass flux at  $t = 10.0\mathrm{\ days}$ found in model M300-9.

In Fig.~\ref{fig.f300-5}, we show the radial profile of the ratio of the mass taken away by the wind to the mass of the injected gas over the $499.89\mathrm{\ days}$ (see Eq. 17). The value of $f_{\rm wind}$ increases with radius within $r \lesssim 1400R_{\rm s} $. The maximum value is approximately $75\%$. It is consistent with the result above that roughly $25\%$ of the injected debris is accreted by the black hole. The reason for the increase of $f_{\rm wind}$ with radius inside $r \lesssim 1400R_{\rm s} $ is the same as we stated before. The value of $f_{\rm wind}$ in model M300-5 is slightly higher than that in model M300-9. Outside $5\times10^{4} R_{\rm s}$, the value decreases with radius. The reason is the same as that stated above in Model 300-9.

In Fig.~\ref{fig.ep300-5}, we plot the radial profiles of the kinetic power carried by the winds at different snapshots for model M300-5. It also shows that the winds are powerful, with a maximum kinetic power reaching $4\times10^2L_{\rm Edd}$ that is lower than that in model M300-9, but still has a magnitude of $10^{46} {\rm erg\cdot s^{-1}}$ at $t = 10.0\mathrm{\ days}$.

The main difference between the two models occurs in the angular distributions of mass flux and kinetic power of wind. In order to illustrate this point, we first introduce the Bernoulli parameter. Fig.~\ref{fig.be300-5} shows the Bernoulli parameter calculated in the units with $G = M_{\rm BH} = R_{\rm s} = 1$ at four snapshots. The Bernoulli parameter generally increases with decreasing $\theta$ at a fixed radius. Also, at a fixed $\theta$, the Bernoulli parameter increases with increasing radius. However, unlike the results shown in Fig.~\ref{fig.be300-9}  for model M300-9, in the region close to the mid-plane, model M300-5 develops a larger area in which the Bernoulli parameter is negative. The reason for the negative Bernoulli parameter is that the radiation enthalpy is much lower.

The change of metallicity affects the system in different channels and we attribute this phenomenon mainly to the difference in the accretion rate. The lower accretion rate reduces the amount of released orbital energy, which makes the radiation weaker.

For $t < 200$ day, the mass accretion rate in model M300-5 is lower than that in model M300-9. The reasons are as follows: First, the increase in metallicity decreases the maximum injection rate, which directly makes the accretion rate in the model M300-5 lower than that in the model M300-9. Second, the injection radius $R_{\rm C}$ in model M300-5 is larger, which is  $R_{\rm C} = 79.44R_{\rm S}$. Some of the injected mass stays within $34.68R_{\rm S} $--$ 79.44R_{\rm S}$ and part of the gas is launched as winds. The mass flux that actually flows into the area $r<34.68R_{\rm S}$ is much lower than the injection rate. The lower accretion rate in model M300-5 makes lower radiative efficiency. Thus, the radiation energy in model M300-5 is relatively lower. Therefore, the radiation enthalpy in model M300-5 is lower, which makes the Bernoulli parameter (especially in the region around the mid-plane) smaller in this model.

\begin{figure}[htbp]
\centering 
\includegraphics[width=0.9\textwidth]{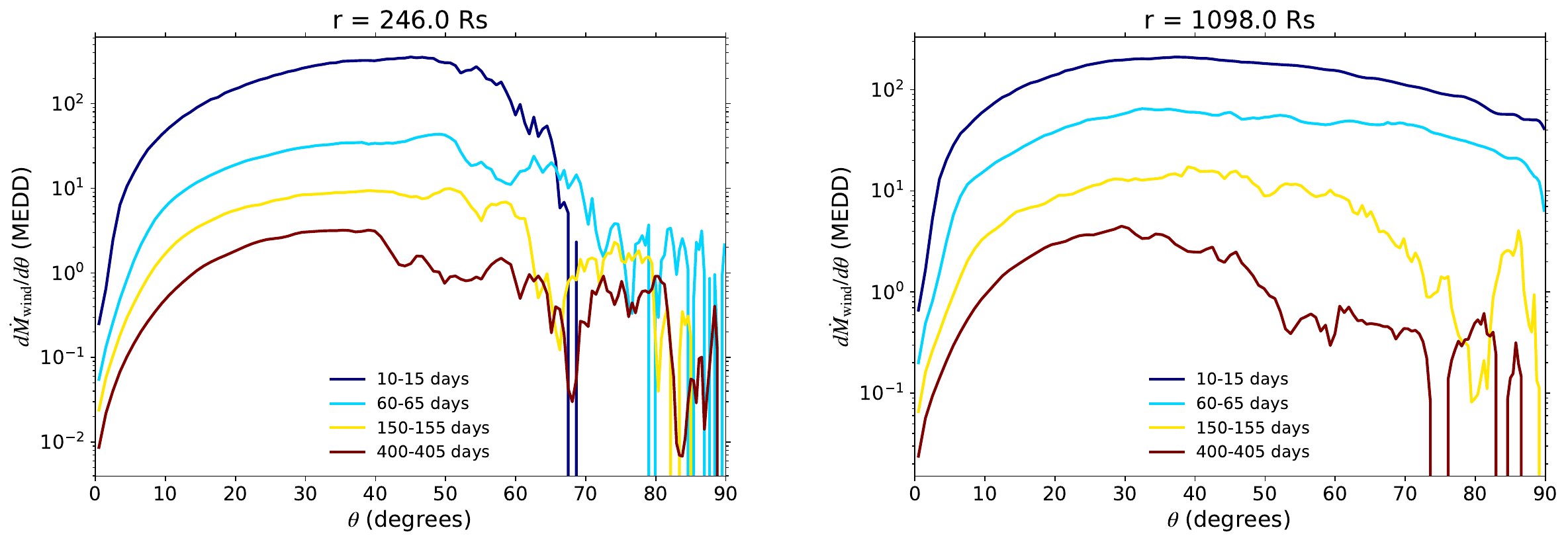}

\caption{
    The angular($\theta$) distribution of mass flux of wind in unit of Eddington accretion rate for model M300-5 at $r = 246\mathrm{R_s}$ (left panel) and $r = 1098\mathrm{R_s}$ (right panel). In order to eliminate the fluctuation, we do time average to the wind mass flux. The blue, cyan, yellow and red lines represent average period of 10-15, 60-65, 150-155 and 400-405 days respectively.
}
\label{fig.vt300-5}
\end{figure}

\begin{figure}[htbp]
\centering  
\includegraphics[width=0.9\textwidth]{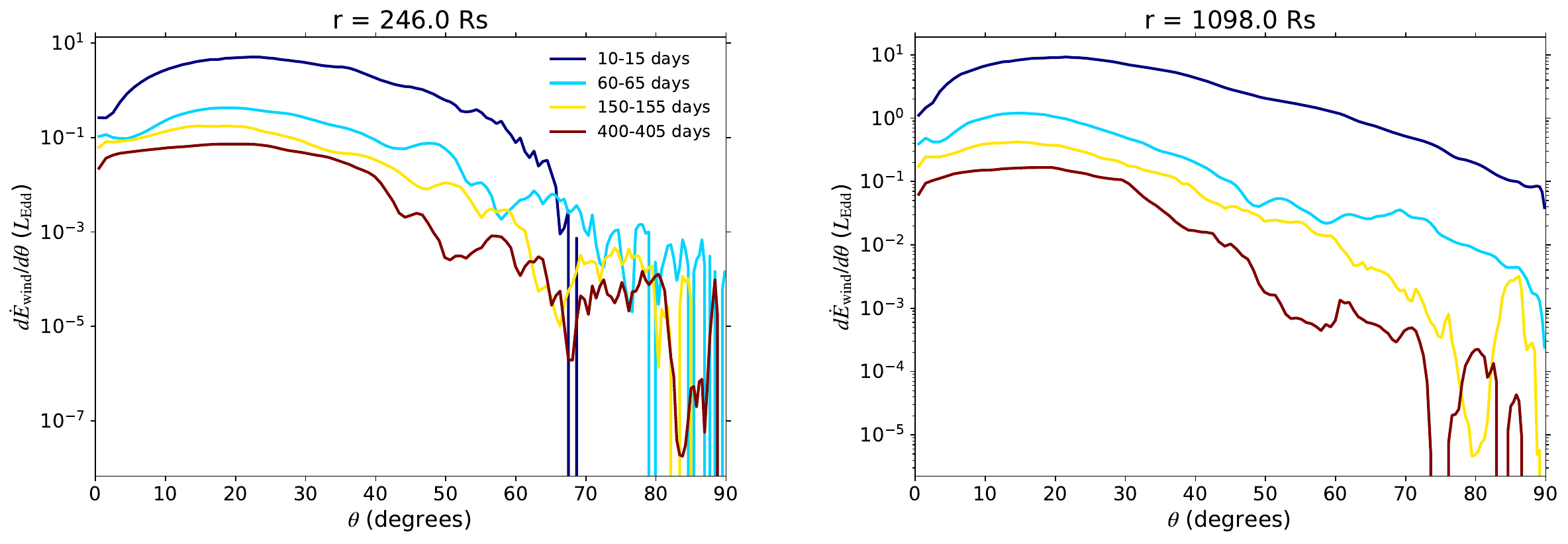}
\caption{
    The angular($\theta$) distribution of kinetic power of wind in unit of $L_{\rm Edd}$ for model M300-5 at $r = 246.0\mathrm{R_{\rm s}}$ (left panel) and $r = 1098.0\mathrm{R_{\rm s}}$ (right panel). In order to eliminate the fluctuation, we do time average to the kinetic power. T average period of 10-15, 60-65, 150-155 and 400-405 days respectively.
}
\label{fig.et300-5}
\end{figure}

To present the results quantitatively, we plot Fig.~\ref{fig.vt300-5} to show the angular distribution of wind mass flux at two different locations, $r = 246R_{\rm s}$ and $r = 1098R_{\rm s}$. To eliminate fluctuations, we time-averaged the results over 5-day periods around different time snapshots. In the region $\theta \lesssim40^\circ$, the mass flux increases steeply between $\sim0^\circ $and$ 10^\circ$ and then increases slowly as $\theta$ increases, similar to the results of model M300-9 (see Fig.~\ref{fig.wt300-9}). A sharp decline in mass flux and extreme fluctuation occur in the region $\theta \gtrsim 70^\circ$ at an early time at $r = 246R_{\rm s}$. At some $\theta$ angles, there is no measurable wind due to $\mathrm{Be} < 0$. However, as mentioned above, after $\sim 200\rm\ days$, the accretion rate of model M300-5 becomes comparable and even exceeds that of model M300-9. As shown in the left panel of Fig.~\ref{fig.vt300-5}, the angular distribution of the mass flux becomes smooth (although still fluctuating) at a late time.
The $\rm Be$ is positive at larger radii $r = 1098R_{\rm s}$ at an early time, and the wind flowing to these areas has enough energy to move further. The angular distribution of the mass flux is smooth. At $r = 1098R_{\rm s}$, the mass flux only shows a same decline and fluctuation feature at late times (right panel of Fig.~\ref{fig.vt300-5}) when the negative $\rm Be$ region expands to $\gtrsim 1000R_{\rm s}$ (see Fig.~\ref{fig.be300-5}). The results show that, at large radii (e.g., $1098R_{\rm s}$), the angular distribution of the properties of the wind is quite similar in the two models. However, in the inner region (e.g, $246R_{\rm s}$), the angular distribution of the properties of the wind is quite different in the two models at early time.

We also plot the angular distribution of the kinetic power carried by the wind in Fig.~\ref{fig.et300-5} and find the same phenomenon. Similar to the results of Fig.~\ref{fig.et300-9}, the kinetic power increases with $\theta$ and peaks at $\theta \sim 20^\circ$. Then it decreases with $\theta$ between $20^\circ $ and $70^\circ$. Beyond $70^\circ$, the kinetic power fluctuates with $\theta$. The velocity shown in Fig.~\ref{fig:vtheta300-5} is similar to that in Fig.~\ref{fig:vtheta300-9} for model M300-9, so the fluctuation is due to the fluctuation of the wind mass flux (see Fig.~\ref{fig.vt300-5}).

\begin{figure}[htbp]
\centering  
\includegraphics[width=0.9\textwidth]{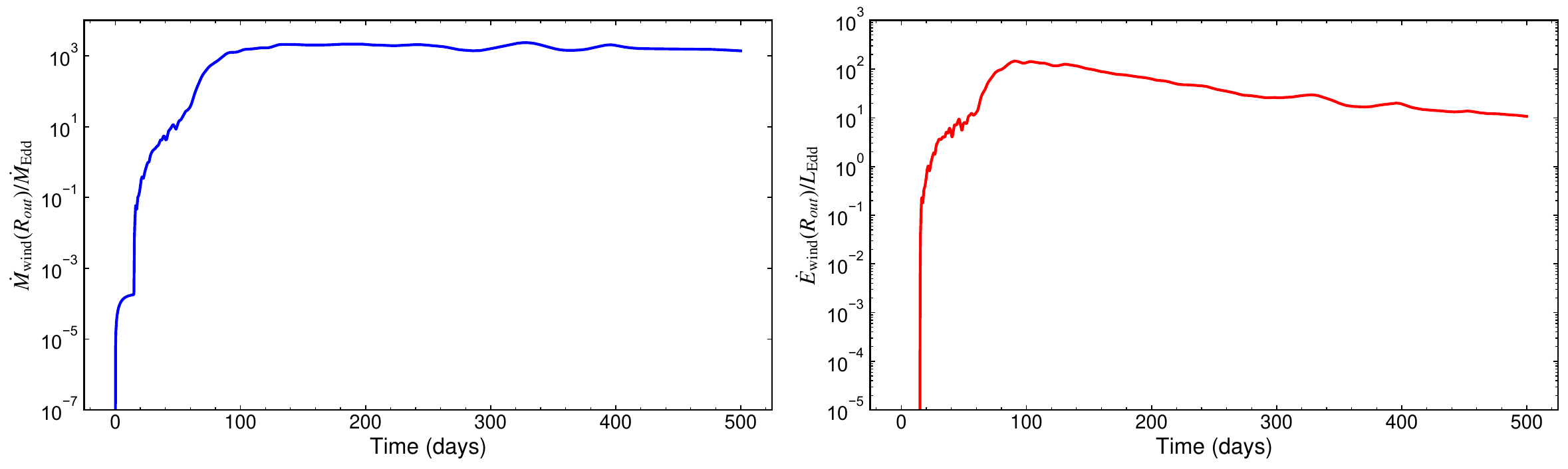}
\caption{
The temporal profile of the mass flux (left panel) and kinetic power (right panel) of wind for model M300-5 measured at the simulation
outer boundary $10^5R_{\rm }$.}
\label{fig.es300-5}
\end{figure}

 Fig.~\ref{fig.es300-5} shows the time evolution of the mass flux and the kinetic power of the wind at the outer radial boundary $10^5R_{\rm s}$ for model M300-5. It shows a similar evolution pattern to model M300-9 (see Fig.\ref{fig.es300-9}) after the wind head reaches the boundary at $t\sim18\mathrm{\ days}$. The mass flux increases steeply and reaches a value $1.386\times10^3\dot{M}_{\rm Edd}$ at $t\sim100\ \rm days$ and the kinetic power reaches a peak value $\sim135L_{\rm Edd}$. The peak values of the mass flux and the kinetic power measured at the outer boundary in model M300-5 are lower than those in model M300-9. However, these values at the end of the simulation are $1.384\times10^3\dot{M}_{\rm Edd}$ and $10.7L_{\rm Edd}$, respectively. The values are a little higher in model M300-5. This is because the debris fallback time scale is larger in model M300-5 than that in model M300-9, which makes the system in model M300-5 evolve slower.

\section{Summary and Discussion}

We performed 2D radiation-hydrodynamic simulations to study the accretion-wind system of Pop III star TDEs. In our models, the mass of the disrupted star is $300 M_\odot$. We have two models with  metallicities of $10^{-9} Z_\odot$ and $10^{-5} Z_\odot$, respectively. We assume that after the disruption, debris can be rapidly circularized. We inject the fallback debris at twice the orbital pericenter. We run the simulations to track the TDE accretion system to 500 days. The debris injection rate is still super-Eddington at the end of our simulation.

We find that 35\% and 25\% of the injected debris are ultimately accreted by the black hole in model M300-9 and model M300-5, respectively. The remainder is unbound and expelled in a powerful, radiation-driven wind, the mass flux of which is approximately proportional to the accretion rate. Our results show that the outer boundary of the accretion disk can expand to approximately ten times the injection radius during the simulation. The winds exhibit a strong anisotropy with respect to the viewing angle $\theta$. Velocities decrease from $\sim 0.9c$ near the polar axis ($\theta \sim 0^\circ$) to values smaller than $0.1c$ at the disk plane ($\theta = 90^\circ$). The mass flux increases from minimal values near the poles towards intermediate angles and then decreases with the increase of $\theta$.

Due to the extremely high mass injection rate of Pop III star debris, the density in such a system is so high that a photosphere is created to envelop the whole accretion system. We find that the photosphere is elongated vertically at early times. Then, at late times, it becomes horizontally elongated. In model M300-9, the system is optically thick for observers with any viewing angles $\theta$ at early times $t < 449 $ days.  For model M300-5, we find that the accretion system is optically thick for any viewing angle even at the end of the simulation.

The absence of a funnel in the early time would make the light curve of Pop III star TDEs distinct from normal TDEs. For example, for model M300-9, in the early stages of the event (t $\lesssim$ 449 days), the wind is optically thick for any viewing angles. We have done simple estimations and find that the peak radiation frequency of the blackbody radiation is around $10^{15}-10^{16}$ Hz. After $\sim$ 449 days, the region close to the rotational axis becomes optically thin, and X-rays can leak out there. After considering the cosmic redshift (for example, z=10), we expect to observe a black body spectrum with Infrared/Optical peak frequency at early stages. After $\sim449$ days, in addition to the infrared/Optical emission, we expect to also observe UV emission (redshifted X-ray emission). In addition, the wind can interact with CNM, which may generate shocks, accelerate power-law electrons, and produce radio emission via synchrotron process. In our forthcoming paper, we will introduce the details of the radiation properties of the Pop III star TDEs.

Finally, we present the properties of the wind that escaped from our simulation domain during 500 days, including the evolution of mass flux and kinetic power at $r = 10^5R_{\rm s}$. For model M300-9, the wind head arrives at the outer boundary at $t\sim15\mathrm{\ days}$, after which the value of mass flux and kinetic power increase rapidly. At $t \sim 100\mathrm{\ days}$, the mass flux reaches a peak value of $2.64\times10^3\dot{M}_{\rm Edd}$ and slightly decreases to a value of $6.08\times10^2\dot{M}_{\rm Edd}$ at $t = 499.89$ days. At the same time, the kinetic power could reach a maximum value of $218L_{\rm Edd}$, decreasing with time to $3.86L_{\rm Edd}$. For model M300-5, we find that the wind reaches the outer boundary at $t\sim18\mathrm{\ days}$ and reaches a value of $1.386\times 10^3\dot{M}_{\rm Edd}$ for mass flux and $135L_{\rm Edd}$ for kinetic power at $100$ days. The evolution of escaped wind in model M300-5 is similar to that in model M300-9 but proceeds slower due to a larger fallback time scale. At $t = 500$ days, the mass flux and kinetic power are $1.384\times10^3\dot{M}_{\rm Edd}$ and $10.7L_\mathrm{\rm Edd}$, respectively, for model M300-5.

In our simulations, we assume that `circularization' of fallback debris is efficient. However, the efficiency of debris circularization remains debated \cite{Kochanek1994}; \cite{Bonnerot2016}; \cite{Hayasaki&Stone&Loeb2016}; \cite{Bonnerot&Rossi&Lodato2017}; \cite{Bonnerot&Lu2020}; \cite{Rossi&Servim&Kesden2021}. Non-circularized debris would retain a higher mechanical energy, potentially generating stronger winds than those predicted in our models. In the future, it is necessary to review the wind from Pop III star TDEs in which the fallback debris is not circularized.

In our simulations, we use a viscous stress tensor to mimic the angular momentum transfer of the accretion flow. In reality, a magnetic field should be present and responsible for the angular momentum transfer of the circularized accretion flow (\cite{1991ApJ...376..214B}). In addition, in the presence of a magnetic field, the wind may be driven by the magneto-centrifugal force (e.g., \cite{Blandford&Payne1982}; \cite{2000MNRAS.318..417V}) or magnetic pressure gradient force (e.g.,\cite{1986CaJPh..64..507U}); \cite{1996MNRAS.279..389L}. It is difficult to predict to what extent the results would change in the presence of a strong magnetic field. Therefore, it is quite necessary to review the circularized accretion flow with a magnetic field in the future.

The velocity of the wind close to the rotational axis can be as high as $0.9c$. For such a high speed wind, the special relativistic effects (ignored in this paper) may be important. The properties of the wind may be changed after considering special relativity. We do not know to what extent the properties of the wind would be affected by ignoring this effect. Therefore, it is important to review the wind from Pop III star TDEs with special relativity effects included in the future.

\section{Acknowledgments}

We acknowledge the reviewer for the insightful suggestions that have greatly helped us to present the results better. Y-HS thanks N-DL for his helpful advice on the drawing of the image. D-FB acknowledges support from the National Natural Science Foundation of China (Grants 12173065, 12133008, 12192220, and 12192223) and the China Manned Space Program (Grant CMS-CSST-2021-B02). X-HY is supported by Chongqing Natural Science Foundation (grant CSTB2023NSCQ-MSX0093) and the Natural Science Foundation of China (grant 12347101). LC is supported by NSFC (12173066), National Key R$\&$D program of China (2024YFA1611403), National SKA Program of China (2022SKA0120102) and Shanghai Pilot Programme for Basic Research, CAS Shanghai Branch (JCYJ-SHFY-2021-013). Computational work was performed using high-performance computing resources at the Advanced Research Computing Core Facility of Shanghai Astronomical Observatory.


\bibliography{sample701}{}
\bibliographystyle{aasjournalv7}



\end{document}